\documentclass[aps,prc,twocolumn]{revtex4}
\usepackage{graphicx}
\usepackage{amsmath,bbm}
\usepackage{amssymb,bm}
\usepackage{array}
\usepackage{multirow}
\usepackage{slashed}
\usepackage{caption}
\usepackage{subcaption}
\newcommand\nd{\noindent}

\newcommand \sL {\mathcal{L}}

\newcommand \sV {\mathcal{V}}
\newcommand \bx {{\bf{x} }}
\newcommand \bp {{\bf{p} }}

\newcommand \sE {\mathcal{E}}

\begin{document}
\title{Entropy considerations in Many-Body Gravity and General Relativity, and the impact on cosmic inflation}
\author{S.~Ganesh\footnote{Corresponding author:\\Email: gans.phy@gmail.com}}
\affiliation{Independent Researcher}
\begin{abstract}
	Many body gravity (MBG) is a novel modified theory of gravity formulated in a 5-D space-time-temperature framework, in which the variation in temperature is recast as a variation in the 5-D metric.
	Previous work on MBG has shown that it can reproduce galaxy rotation curves, radial acceleration relation and the weak gravitational lensing of the bullet cluster, without the inclusion of dark matter.
In this work we show that MBG can reproduce cosmic inflation, and in the process, analyze fundamental relations between interaction, time and gravity. 
To analyze cosmic inflation using interacting massless scalar fields, we first analyze theoretically a hypothetical universe with a single massive particle, or a collection of non-interacting massive particles. 
A quantitative relation between time and interaction is developed using Quantum Field Theory (QFT), which suggests that the notion of time becomes ill-defined for such a universe.
The mass terms in MBG and General Relativity cause a discrepancy with the QFT results.
An interacting massless scalar field then becomes a necessity to resolve the issue at the onset of inflation.
However, the entropic terms in the MBG field equations are seen to be consistent with the QFT results and further accelerate inflation. 
The slow-roll condition is shown to be a natural consequence of the Euler-Lagrange equations of motion governing the massless scalar field in 5-D space-time-temperature, during the early phase of inflation.
Finally, the MBG field equations are solved in the context of a Friedmann metric, leading to inflation. The matter era is also investigated. 
\vskip 0.5cm

{\nd \it Keywords } : Many body gravity , Friedmann universe, dark matter, Modified gravity, QFT, Cosmic Inflation \\
{\nd \it PACS numbers } : 04.50.Kd,  11.10.Wx, 11.10.-z,

\end{abstract}

\maketitle
\section{Introduction}
\label{sec:intro}
The theory of gravity proposed by Einstein~\cite{Ein1} has been extremely successful in explaining several phenomena associated with gravity.
	Various experiments~\cite{shap1, hol, sch, foma, shap2,ber}
	including the detection of gravitational waves~\cite{ligo1, ligo2} and the gravitational lensing~\cite{lens1, lens2} have shown that Einstein's general theory of relativity is correct.
	A notable exception has however been the galaxy rotation curves.
	Newtonian gravity, or Einstein's general theory of relativity, predicts that the orbital speeds of stars around the center of the galaxy to decrease radially as one goes away from the center. 
But, the rotational speed of the stars is observed to be approximately constant~\cite{rub1, rub2, galaxy1, galaxy2}.
	The theories propounded to explain the deviations from Einstein's gravity can be classified into two categories. The first category hypothesizes the existence of a hidden mass or dark matter~\cite{dark1, dark2, dark3, dark4, dark5}, while the second examines the modification of the laws of gravity, especially for weak gravitational fields~\cite{mond1, mond2, mond3, mond4, mond5, entropic}.
MBG is a novel modified theory of gravity, which unfolded from attempts to model spatial and temporal variations in temperature.
The galaxy rotation curves, the RAR~\cite{gans8, gans8E}, and the weak gravitational lensing of the Bullet cluster~\cite{gans9} can be explained by MBG.
In the bullet cluster system, most of the baryonic mass is located in a different region than predicted by weak gravitational lensing 
~\cite{bullet1, bullet2, bullet3, bullet4}.
This is easily explained by the presence of dark matter in the relevant regions.
	This simultaneously makes the bullet cluster a smoking gun evidence for dark matter, and a challenge for alternative gravity theories to explain. 
MBG takes into account the variation in the 3-D hydrogen gas density around the galaxy clusters. The variation in the 3-D density leads to entropy variations, which manifest in space-time curvature due to the additional thermal (entropic) terms in MBG~\cite{gans9}. 
In the case of galaxy rotation curves, the interaction between the stars and the gas within a galaxy leads to entropy variations, which is captured by MBG~\cite{gans8, gans8E}.
To summarize, MBG is able to model galaxy rotation curves and weak gravitational lensing without dark matter, since it also captures variations in entropy along with mass.
The foundation of MBG's theory is rooted in the modeling of spatial and temporal variations of thermal systems, which can be confirmed experimentally~\cite{gans7}. Additional background of the development of the theory can be found in ~\cite{gans5, gans6}. Moreover, the theory is manifestly relativistic.

Cosmic inflation is an important cosmological theory, which can explain the lack of magnetic monopoles, the horizon problem, etc.~\citep{alan}.
If there is a negative pressure, General Relativity predicts cosmic inflation within the Friedmann metric framework.
A massless scalar field can produce negative pressure if there is a slow-roll, which means that the time derivative of the scalar field is zero.
Any modified theory of gravity is expected to reproduce the cosmic inflation, or provide another mechanism to explain the lack of magnetic monopoles, horizon problem, etc.
In this work, it is shown that solving the MBG equations, using the Friedmann-Lemaitre-Robertson-Walker (FLRW) metric, does lead to cosmic inflation as a possible outcome.
Only a single field cosmic inflation is considered.
Apart from negative pressure, we investigate if there are further fundamental reasons for requiring an interacting massless scalar field. We do this by demonstrating that the opposite case of non-interacting massive particles leads to a disparity between the results from QFT and General Relativity.
We also show that the entropic terms in the MBG field equations enable a quantitative relation between interaction, gravity, and time, which is consistent with the results of QFT and accelerates the cosmic inflation.
The entropic terms are the equivalent of dark matter in the MBG theory. 


To this end, we first probe the MBG theory and General Relativity from the fundamental viewpoint of time and particle interactions.
Leaving aside quantum gravity, we inquire if there is any fundamental element that must be included in a theory of gravity.
To explore this aspect, we consider the following cases as a first intuition.

\begin{itemize}
\item Case I: 
Consider a hypothetical universe that consists of only one massive particle.
The notion of time becomes ambiguous or ill-defined.
However, Einstein's theory suggests that a single massive particle could cause a space-time curvature, according to the equations of General Relativity.
This leads to a disparity, which is investigated in the context of both General Relativity and MBG in Sec.~\ref{sec:single_particle_MBG}.

\item Case II: We now promote the hypothetical universe to a collection of stationary, non-interacting massive particles.
The absence of interaction results in the absence of motion.
Again, the concept of time becomes ambiguous. 
The relation between time and interaction is formally derived using QFT in Sec.~\ref{sec:interaction}.
An explanation is required for the absence of interaction, as two masses would always have gravitational interaction.
To make them non-interacting, we could add a minuscule charge in each of the mass, so that the electrostatic repulsion balances the gravitational attraction. The net force is zero. 
Another thought experiment could involve using $W^{-}$ bosons and decreasing the electric coupling until the gravitational force and electric force are equal.


\item Case III: In the third step, we introduce interactions, i.e., the coupling constant, $\lambda$, between them is non-zero.
The particles start moving in each other's influence. Their position and velocity change with time. Thus, time itself gets a meaning, and it flows.
There will be some subtleties involving the gravitational coupling vs. the coupling for the other fundamental forces, which we will delve into later in Sec.~\ref{sec:single_particle_MBG}.
For the sake of simplicity, let's ignore the gravitational coupling for now because it's very weak.

\item Case IV: In this final step, we divide the universe into multiple regions, as shown in Fig.~\ref{fig:mbg_regions}. In each region, the scenario is the same as case III. The particles move under each other's influence. 
However, the coupling constant $\lambda_i$ differs from region to region,
i.e., $\lambda_1 \ne \lambda_2 \ne \lambda_3 ...$. Within a region, $\lambda$ is taken as a constant. 
The progression of time should be different in each region.
Thus, by adjusting the interaction rate across space, time in different regions can be altered.
Time variation across space leads to space-time curvature.
\begin{figure}
\includegraphics[width = 80mm,height = 80mm]{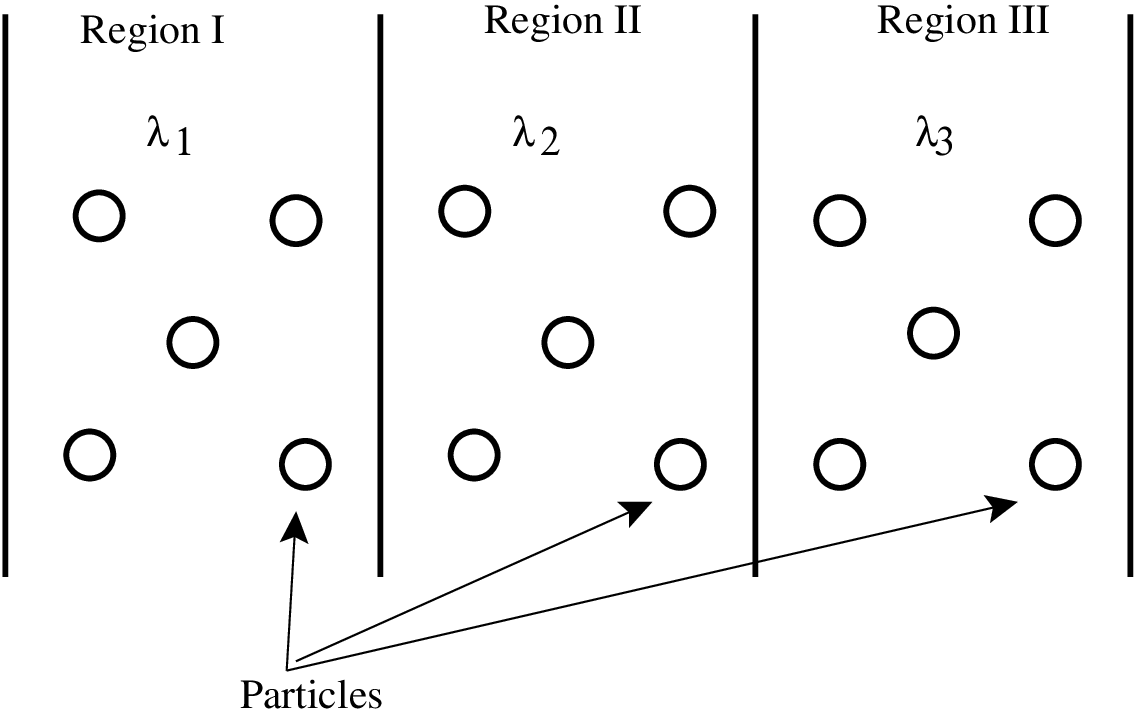}
        \caption{ Different regions in space with different coupling constants, $\lambda$.}
\label{fig:mbg_regions}
\end{figure}
\end{itemize}

Cases I - IV show that interaction and the concept of time are intricately interwoven.
If such is the case, then the laws of gravity must incorporate interaction. 
Equivalently, entropy being a manifestation of interaction, should be considered as an ingredient in the laws of gravity.
Within the context of this work, the terms entropy and interaction are used interchangeably.
The topic of interaction and coupling constant is treated quantitatively, in Sec.~\ref{sec:interaction} and~\ref{sec:single_particle_MBG}.
The MBG theory is a "many body" theory, and it intrinsically incorporates interaction or entropy. However, in the case of General Relativity, it must be inserted externally.

These considerations become significant at the beginning of time, namely during the start of cosmic inflation.
The presence of mass can prevent taking the limit, time $\rightarrow 0$, due to the anomaly discussed in cases I and II. 
The need to eliminate the problematic mass term necessitates the consideration of a massless scalar field, $\hat{\phi}$, during inflation.
A massless non-interacting scalar field leads to a null stress energy tensor. Hence, literature introduces interaction in terms of an external potential $V(\hat{\phi})$ \citep{potential1, potential2}, while working with Einstein's gravity.
MBG inherently contains interaction (entropic) terms in its formulation, and hence an explicit introduction of $V(\hat{\phi})$ is not required.
MBG is essentially Einstein's equations in 5-D space-time-temperature. Hence, the massless scalar field is analyzed within the framework of a field theory in 5-D space-time-temperature, developed in Ref.~\cite{gans7}.
The slow roll condition is seen as a natural consequence of the Euler-Lagrange equations of motion in 5-D at the onset of inflation. 
Subsequently, inflation is analyzed using MBG.
Finally, the MBG field equations are solved within the framework of the Friedmann metric, resulting in an inflationary phase. 
In the end, the matter era is also discussed.

We also discuss possible counter arguments to our line of discussion, related to the uncertainty principle or gravitons forming black holes~\cite{graviton}, in Sec.~\ref{sec:counter_arguments}.


	The rest of the article is as follows. 
	An overview of the MBG theory developed in Refs.~\cite{gans8, gans8E} is given in Sec.~\ref{sec:overview}. 
	The relation between time and interaction is examined in Sec.~\ref{sec:interaction} for QED, QCD and the weak force. The Higgs mechanism is also discussed for analyzing mass. 
	This is followed by the discussion on the disparity between QFT and Einstein's gravity in the case of massive single particles or massive particles with zero interactions. 
The compatibility of the entropy terms in MBG theory with the QFT results is also addressed in Sec.~\ref{sec:single_particle_MBG}. 
	A massless scalar particle is shown to support inflation in a 5-D space-time-temperature framework in Sec.~\ref{sec:friedmann}. Furthermore, the Friedmann universe during the inflation and the matter era is also analyzed using the MBG theory in Sec.~\ref{sec:friedmann}.
The resemblance to Newtonian gravity in the weak field limit and at far distances is also elaborated in this section.
	Finally, Sec.~\ref{sec:conclusion} draws the conclusions.


\section{Overview of 5-D thermal field theory and MBG}
\label{sec:overview}
A brief overview of the MBG theory, developed in Ref.~\cite{gans7,gans8} is now provided. For details and interpretation, the reader can refer Refs.~\cite{gans7,gans8}.
Section~\ref{sec:partition} is mostly a summary of Ref.~\cite{gans7}, while,
Sec.~\ref{sec:overview_mbg}, primarily summarizes Ref.~\cite{gans8}.
The sign convention used for 5-D space-time-temperature would be either $(+,-,-,-,-)$ or $(-,+,+,+,+)$, as specified in the context. The important aspect is that the sign of the metric for the spatial and thermal dimension is identical.
\subsection{The 5-D thermal field theory }
\label{sec:partition}
In Ref.~\cite{gans6}, the Quantum field theoretic aspects of a non-time varying thermal field with spatial variations, were developed. In Ref.~\cite{gans7}, it was extended to time varying thermal fields in the 5-D space-time-temperature $(t,x,y,z,i\beta)$.
We now summarize the Quantum field theoretic aspects of the more generic case of a time varying thermal field, developed in Ref.~\cite{gans7}. 
The field theory is developed for a Lagrangian in 5-D space-time-temperature. The Lagrangian density in a 5-D space, for a neutral scalar field would be,
\begin{equation}
\label{eq:lagrangian5D}
        \sL(\hat{\phi},\partial_a \hat{\phi}) =  \frac{1}{2} \left ( \partial_{a}\hat{\phi} \partial^{a}\hat{\phi} - m^2\hat{\phi}^2 \right ),
\end{equation}
with the index $a=0,1,2,3,4$ corresponding to the dimensions $[t,x,y,z, \tau]$.
$\tau$ is the inverse temperature and varies from $0$ to $\beta$.
The sign convention used is (+,-,-,-,-).
This gives rise to the equation of motion:
\begin{equation}
        \partial_a\partial^a\hat{\phi} + m^2\hat{\phi} = 0.
\end{equation}
The constraining 5 momentum delta function would be, 
\begin{equation}
	\label{eq:delta5D}
	\delta( E^2 - \omega^2 - \bp^2 - m^2),
\end{equation}
and the 5-D integral measure is:
\begin{multline}
        \label{eq:invariant}
        \frac{1}{\beta}\sum_{n}\int \frac{d^3p}{(2\pi)^3}\frac{dE}{2\pi} 2\pi\delta(E^2 - \omega_n^2 - \bp^2 -m^2 )\\
        =  \frac{1}{\beta}\sum_{n}\int \frac{d^3p}{(2\pi)^3} \frac{1}{2E},
\end{multline}
where, the Matsubara frequency, $\omega_n = \frac{2n\pi}{\beta}$, for a boson.
As mentioned in Ref.~\cite{gans6}, $E$, may be considered the original intrinsic energy of a particle, and $\omega_n = iE_c$, can be considered as the interaction energy of the particle with the thermal medium. The variable, $\omega_n$, determines the decay or enhancement of a particle wave-function with temperature (for example, the Dirac spinor in Ref.~\cite{gans6}).
Since $E$ and $\omega_n$ lie in orthogonal dimensions (conjugate momenta to time and temperature respectively), the magnitude of the total energy is then 
\begin{equation}
	\label{eq:Emag5D}
= \sqrt{E^2 - \omega_n^2}.
\end{equation}
It is intuitive, that a particle's 3-momentum would be affected by both $E$ and $\omega_n$, and not just $E$. 
Thus, one may consider 
\begin{equation}
	\label{eq:Eomega}
E^2 - \omega_n^2 = \bp^2 + m^2. 
\end{equation}
A portion of the particle's original energy, $E$, is lost due to interaction with the thermal medium. This provides an intuition behind the delta function, $\delta(E^2 - \omega_n^2 - \bp^2 - m^2)$.

The operator for a neutral scalar field in 5-D space-time-temperature is,
\begin{multline}
        \label{eq:operator5D}
        \hat{\phi}(\bx,\tau,t) = \frac{1}{\beta}\sum_n \int \frac{d^3p}{(2\pi)^3} \frac{1}{\sqrt{2E_p}}\\
      \times  \sum_s \left ( a^{\dagger}_{\bp,\omega_n} e^{-ipx}e^{-i\omega_n \tau} + a_{\bp,\omega_n}e^{ipx}e^{i\omega_n \tau} \right ).
\end{multline}

The operator, $a^{\dagger}_{\bp,\omega_n}$, creates a particle with 3-momentum $\bp$, and Matsubara frequency $\omega_n$.

One may premise the below commutation relation:
\begin{equation}
        \label{eq:commutation5D}
        [a_{\bp_1,\omega_{n1}}, a^{\dagger}_{\bp_2,\omega_{n2}}] = (2\pi)^3\delta^3(\bp_1 - \bp_2) \zeta(\beta)\delta_{n1,n2},
\end{equation}
where, $\zeta(\beta)$ is a scalar normalization function, and needs to be determined.
Let us define,
\begin{eqnarray}
        \label{eq:pcreation}
        \nonumber       a_{\bp} = \sum_n f(\omega_n) a_{\bp,\omega_n}, \\
        a^{\dagger}_{\bp} = \sum_n f^*(\omega_n) a^{\dagger}_{\bp,\omega_n}.
\end{eqnarray}
Eq.~\ref{eq:pcreation} can be interpreted in the following way. When a momentum state $|\bp\rangle$ is created, then $|\bp\rangle$ itself can be treated as a superposition of the momentum-Matsubara eigenstates $|\bp,\omega_n\rangle$, with probability amplitudes $f(\omega_n)$.
Since $f(\omega_n)$ is a probability amplitude, $\sum_n|f(\omega_n)|^2 = 1$.
In Ref.~\cite{gans7}, it was shown that the commutator,
\begin{multline}
\label{eq:equaltau}
[a_{\bp_1}, a^{\dagger}_{\bp_2}] 
        = \sum_{n1} (2\pi)^3 \zeta(\beta) \delta^3(\bp_1 - \bp_2)|f(\omega_{n1})|^2.
\end{multline}
Since $\sum_{n1} |f(\omega_{n1})|^2 = 1$,  let us assign $\zeta(\beta) =  1$, in Eq.~\ref{eq:equaltau}, to obtain,
\begin{equation}
        \label{eq:commutator4D}
        [a_{\bp_1}, a^{\dagger}_{\bp_2}] = (2\pi)^3\delta^3(\bp_1 - \bp_2).
\end{equation}
Thus, the usual commutation relation between the 3-momentum annihilation and creation operator is recovered.

\subsection{An overview of the MBG theory}
\label{sec:overview_mbg}
A brief overview of the MBG formulation developed in Ref.~\cite{gans8}, is now presented.
In literature, index 0 is usually reserved for the time dimension. In order to be more compatible with this convention, We modify the convention for indices used in ~\cite{gans7, gans8, gans8E, gans9}, where, the $0^{th}$ index was used for temperature. The $0^{th}$ dimension is now used for time, and the last dimension for temperature.
We use the letters, $a$ and $b$, as indices for the 5-D space-time, i.e., $a$, $b$ = 0, 1, 2, 3, 4, with the index 0 referring to the time dimension, and the index 4 referring to the temperature dimension,
The letters, $\mu$ and $\nu$, are indices for the 4-D Lorentzian space-time, i.e., $\mu$, $\nu$ = 0, 1, 2, 3.
A superscript, $(N)$, within brackets, refers to $N$ dimensional space. For example, $\nabla^{(4)}_{\mu}$, refers to the covariant derivative in 4-D space-time.

Let $\beta({\bf x},t)$ be the inverse temperature at space-time $({\bf x},t)$. Then, we represent $\beta({\bf x},t)$ as:
\begin{equation}
	\label{eq:define_s}
	\beta({\bf x},t) = s({\bf x},t)\beta_c,
\end{equation}
where, $\beta_c$ is now a constant. All the variation in $\beta({\bf x},t)$ is now captured in $s({\bf x},t)$
	Based on the theory developed in Ref.~\cite{gans7}, we consider the 5-D metric, $g^{(5)}_{ab}$, as:
            \begin{equation}
		    \label{eq:5Dmetric}
		    g^{(5)}_{ab} = 
                \left[ \begin{array}{c c}
			g^{(4)}_{\mu\nu} & 0\\
			0 & s^2(\bx,t)\\
                \end{array} \right ],
            \end{equation}
where, $g^{(4)}_{\mu\nu}$, is the usual metric tensor in 4-D space-time, and is purely due to gravitational fields, while, $s(\bx,t)$ captures the variation in the inverse temperature, $\beta$.
The sign convention used is (-,+,+,+,+).
	For a system in thermal equilibrium, the Einstein field equation in 5-D space-time-temperature is given by~\citep{gans7}:  
\begin{equation}
	\label{eq:EFE5D}
	R^{(5)}_{ab} - \frac{1}{2}g^{(5)}_{ab} R^{(5)} = \frac{8\pi G}{c^4} T^{1(5)}_{ab},
\end{equation}
where, the stress energy tensor, $T^{1(5)}_{ab}$, is given by:
\begin{equation}
	T^{1(5)}_{ab} = (\rho + \frac{P_1}{c^2})u_a u_b + P_1g^{(5)}_{ab},
\end{equation}
	with, $\rho$ being the density, and $P_1$, the pressure.
The Ricci tensor, $R^{(5)}_{ab}$, can be expressed in terms of the 4-D covariant derivative operator, $\nabla^{(4)}_{\mu}$, and the 4-D Ricci tensor, $R^{(4)}_{\mu\nu}$, as:
            \begin{equation}
		    \label{eq:ricci5D}
		    R^{(5)}_{ab} = 
                \left[ \begin{array}{c c}
			R^{(4)}_{\mu\nu} - \frac{1}{s}\nabla^{(4)}_{\mu} \nabla^{(4)}_{\nu} s & 0\\
			0 & R_{\beta\beta} \\
                \end{array} \right ],
            \end{equation}
where, 
	$ R_{\beta\beta} = -s\nabla^{(4)\mu} \nabla^{(4)}_{\mu} s $.
	In contrast, for systems that are non-interacting and in complete non-equilibrium, the particles behave as if there are no other particles. 
Without an ensemble or temperature concept, the temperature dimension becomes meaningless. 
To model this scenario, let us take the thermal gradient in the zero limit, i.e., $\partial_{\mu}s \rightarrow 0$, followed by $s= 0$. 
A system with zero temperature cannot have thermal gradients. This mandates $\partial_{\mu}s \rightarrow 0$. 
Subsequently, assigning $s=0$, eliminates the temperature dimension from the metric $dS^2 = s^2d\beta^2 - dt^2 + dx^2 + dy^2 + dz^2$, leading to a 4-D space-time.
As, $\partial_{\mu}s \rightarrow 0$, it is evident that the 5-D Ricci tensor in Eq.~\ref{eq:ricci5D}, is reduced to a 4-D Ricci tensor. The reduction to 4-D field equations is explained in more detail in Ref.~\cite{gans7, gans8}. 
One may also refer to Ref.~\cite{gans7} for the relations between the field theories in 5-D and 4-D.
	After reduction to 4-D, the 4-D Einstein's field equations, represented in 5-D, are:
            \begin{equation}
	\label{eq:EFE4D}
                \left[ \begin{array}{c c}
			R^{(4)}_{\mu\nu} & 0\\
			0 & 0\\
                \end{array} \right ]
		    -\frac{1}{2} \left[ \begin{array}{c c}
				     g_{\mu\nu} R^{(4)} & 0\\
			0 & 0\\
                \end{array} \right ]
   		    = \frac{8\pi G}{c^4} \left[ \begin{array}{c c}
			    T^2_{\mu\nu} & 0\\
			0 & 0\\
                \end{array} \right ].
            \end{equation}
		    The stress energy tensors, $T^1$ and $T^2$, have the same $\rho$, but different pressures, $P_1$ and $P_2$. $P_1$ is the pressure due to an ensemble interacting gravitation-ally, while $P_2 \sim 0$ in the absence of any ensemble (a single particle has no concept of pressure).

	                An equation representing the behavior of a partially thermalized system, needs to be a generalization of Eqs.~\ref{eq:EFE5D} and~\ref{eq:EFE4D}, with  Eqs.~\ref{eq:EFE5D} and~\ref{eq:EFE4D} being special cases. 
			To motivate the generalization, let us rewrite Eq.~\ref{eq:EFE5D} by taking all terms that depend on $s$ to the R.H.S.. This is motivated by the fact that $s$ is related to inverse temperature, and thus related to inverse energy, and consequently, can be interpreted to act as a source causing space-time curvature.
		    We also apply the simplification that, for a time-invariant system, $u^0 = c\frac{\partial \beta}{\partial \tau_p} = 0$, where $\beta$ is the inverse temperature, and $\tau_p$ is the proper time. In terms of 4-D operators, Eq.~\ref{eq:EFE5D} then becomes,
\begin{multline}
	\label{eq:EFE5D_2}
                \left[ \begin{array}{c c}
				    R^{(4)}_{\mu\nu}  & 0\\
			0 & 0\\
                \end{array} \right ]
		    -\frac{1}{2} \left[ \begin{array}{c c}
			    g_{\mu\nu} R^{(4)} & 0\\
			0 & 0\\
                \end{array} \right ] = \frac{8\pi G}{c^4} \\
		\times \left[ \begin{array}{c c}
			T^1_{\mu\nu} + \frac{c^4\left (\frac{1}{s}\nabla_{\mu} \nabla_{\nu} s - \frac{g_{\mu\nu}}{s}\nabla^{\alpha} \nabla_{\alpha} s \right )}{8\pi G} & 0\\
			0 & \frac{s^2c^4R^{(4)}}{16\pi G} + s^2P_1 \\
                \end{array} \right ],
\end{multline}
where, $\alpha = 0,1,2,3$, and we have skipped the superscript $(4)$ in $\nabla^{(4)}$ for simplicity of notation. 
The terms in the R.H.S. of Eq.~\ref{eq:EFE5D_2}, can be viewed as new source terms describing a thermalized system, with thermal gradients, which causes curvature of 4-D space-time.
It is now hypothesized that the source term of a partially thermalized system may be considered to be a linear combination of the source terms in Eqs.~\ref{eq:EFE4D} and ~\ref{eq:EFE5D_2}. In other words,
\begin{multline}
	\label{eq:lincomb}
	T^{p(5)}_{ab} = 
		 (1-k) \left[ \begin{array}{c c}
			 T^2_{\mu\nu} & 0\\
			0 & 0\\
                \end{array} \right ] + k \\
		     \times \left[ \begin{array}{c c}
			     T^1_{\mu\nu} + \frac{c^4 \left (\frac{1}{s}\nabla_{\mu} \nabla_{\nu} s - \frac{g_{\mu\nu}}{s}\nabla^{\alpha} \nabla_{\alpha} s \right ) }{8\pi G} & 0\\
			     0 & \frac{s^2c^4R^{(4)}}{16\pi G} + s^2P_1\\
                \end{array} \right ],
\end{multline}
		    where, $k$ represents the degree or extent of equilibration. 
		    Additional interpretation of $k$, can be referred in Ref.~\cite{gans8, gans8E, gans9}.
		    For future use, it is worth noting that one may group the pressure terms together, and define an effective pressure:
\begin{equation}
	\label{eq:Peff}
	P_{eff} = (1-k)P_2 + kP_1.
\end{equation}
The corresponding 5-D Einstein field equations for a partially thermalized system are then:
\begin{equation}
	\label{eq:EFE5D_p}
                \left[ \begin{array}{c c}
			R^{(4)}_{\mu\nu} & 0\\
			0 & 0\\
                \end{array} \right ]
		    -\frac{1}{2} \left[ \begin{array}{c c}
			    g_{\mu\nu} R^{(4)} & 0\\
			0 & 0\\
		    \end{array} \right ] 
		    = \frac{8\pi G}{c^4} T^{p(5)}_{ab},
\end{equation}
with $T^{p(5)}_{ab}$, given in Eq.~\ref{eq:lincomb}.

For very small $k$ (as in a galaxy, or the bullet cluster with hindsight), 
one can neglect factors like $kP_1$, and given that $P_2 \sim 0$,
Eq.~\ref{eq:EFE5D_p} can be simplified to obtain~\cite{gans8}:,
\begin{equation}
\label{eq:mbg_Peq}
	-\frac{1}{2}ks^2R = \frac{8\pi G}{c^4} kP_1 s^2, 
\end{equation}
and,
\begin{multline}
	\label{eq:gal3}
	R_{\mu\nu} - \frac{k}{s}\nabla_{\mu}\nabla_{\nu}s - \frac{1}{2}g_{\mu\nu}R^{(4)} + \frac{kg_{\mu\nu}}{s}\nabla^{\alpha}\nabla_{\alpha}s\\ 
	= \frac{8\pi G}{c^4} \rho u_{\mu}u_{\nu}.
\end{multline}

If $P_2 \approx 0$ (for a non-thermalized single particle), $k$ is very small, 
the Lorentz dilation factor $=\gamma \approx 1$, and 
in the weak field limit, the following relation was derived in Ref.~\cite{gans8}:  
\begin{equation}
	\label{eq:gal8}
	\nabla^2 \phi(\bx) =   4\pi G \rho(\bx) - kc^2\frac{1}{2s(\bx)}\nabla^2s(\bx), 
\end{equation}
where, $\frac{2\phi}{c^2} = g^{(5)}_{00} - \eta^{(5)}_{00}$, and $\eta^{(5)}$ is the metric for a flat 5-D space-time-temperature.
Figuratively, one may say that $4\pi G \rho$ is the energy contribution and the second term, $kc^2\frac{1}{2s(\bx)}\nabla^2s(\bx)$, is the entropy contribution. 
In Ref.~\cite{gans8E}, it was shown that $k$ is negative.
In~\cite{gans8,gans9}, it was shown that for systems like a galaxy, 
\begin{equation}
	\label{eq:gal9}
	s\propto \frac{1}{\phi}. 
\end{equation}
	This gives,
\begin{equation}
	\label{eq:gal10}
	\nabla^2 \phi(\bx) =   4\pi G \rho(\bx) - kc^2\frac{1}{2}\phi\nabla^2\frac{1}{\phi}.
\end{equation}
Equation~\ref{eq:gal10} was the final equation which was used to explain the galaxy rotation curves and the bullet cluster's weak gravitational lensing phenomenon.

\section{Interaction and time}
\label{sec:interaction}
We now explore the relationship between particle interaction and time using QFT. The particle interaction is encapsulated in the form of the coupling constant. 
We will look at the three fundamental forces of quantum electrodynamics (QED), quantum chromodynamics (QCD), the weak force, and finally the Higgs mechanism for generating mass.
Most of the basic QFT formulations are based on Ref.~\cite{PnS}. The basic formulations are modified for the metric below:
\begin{equation}
	\label{eq:g00_metric}
	h_{\mu\nu} = 
                \left[ \begin{array}{c c}
			h_{00} & 0\\
			0 & -\delta_{ij} \\
                \end{array} \right ],
\end{equation}
$\forall i,j = 1,2,3$.
Referring to Fig.~\ref{fig:mbg_regions}, we assume that $h_{00}$ is roughly constant within each of the three regions.
The three regions are sufficiently large, in comparison to the wavelength of the particles, so that $h_{00}$ can be kept constant. We could have considered the metric $h_{00}dt^2 - h_{ij}dx^i dx^j $. 
But, this does not offer any additional insight or generality and only exacerbates the mathematical complexity.
We will show that even this fairly simple treatment with $h_{00}$ as a constant in a given region leads to profound implications. 
It reveals the discrepancy with Einstein's gravity in the case of no interactions.

We first model spinors and relations related to spinors. These would then be useful for both QED and QCD.

\subsection{Spinors}
\label{sec:spinor}
\subsubsection{Dirac Equation }
\label{sec:dirac}
The Dirac equation in curved space time is
\begin{equation}
	\label{eq:Dirac1}
	i\gamma^{\rho} e^{\mu}_{\rho} D_{\mu}\psi - m\psi=0,
\end{equation}
where, $D_{\mu} = \partial_{\mu} - \frac{i}{4}\omega_{\mu}^{\rho\alpha}\sigma_{\rho \alpha} $, $\sigma_{\rho\alpha} = \frac{i}{4}[\gamma_{\rho}, \gamma_{\alpha}]$ and $e^{\rho}_{\mu}$ is the vierbein.
If the metric terms $h_{\mu\nu}$ are almost constants, then $\frac{i}{4}\omega_{\mu}^{\rho\alpha}\sigma_{\rho\alpha} = 0$, The Dirac equation simplifies to  
\begin{equation}
	\label{eq:Dirac2}
	i\gamma^{\rho} e^{\mu}_{\rho}\partial_{\mu}\psi - m\psi \approx 0,
\end{equation}
Let the solution to Eq.~\ref{eq:Dirac2} be 
\begin{equation}
	\psi = a_p u(p) \exp (ih_{\mu\nu}p^{\mu}x^{\mu}),
\end{equation}
where $a_p$ is the annihilation operator.
Substitute $\psi$ in Eq.~\ref{eq:Dirac2}, and
after some algebra, the solution for $u(p)$ is given by,
\begin{equation}
\label{eq:spinor}
		    u{(p)} = 
                \left[ \begin{array}{c }
			\sqrt{h_{\mu\nu}\sigma^{\rho} e_{\rho}^{\nu}p^{\mu}} \zeta \\
			\sqrt{h_{\mu\nu}\overline{\sigma}^{\rho} e_{\rho}^{\nu}p^{\mu}} \zeta \\
                \end{array} \right ]
		=
                \left[ \begin{array}{c }
			\sqrt{p.\sigma } \zeta \\
			\sqrt{p.\overline{\sigma}} \zeta \\
                \end{array} \right ],
\end{equation}
where, $p.\sigma = h_{\mu\nu}\sigma^{\rho} e_{\rho}^{\nu}p^{\mu}$, $\overline{\sigma} = (\sigma^0, -\sigma^1, -\sigma^2, -\sigma^3)$ and $\zeta$ ia a $2\times 1$ spinor.
The spinor product, $\sum_s u^s(p)\overline{u}^s(p)$ is given by:
\begin{equation}
\sum_s u^s(p)\overline{u}^s(p) = 
                \left[ \begin{array}{c c}
			m & {h_{\mu\nu}\sigma^l e_l^{\nu}p^{\mu}} \\
			{h_{\mu\nu}\overline{\sigma}^l e_l^{\nu}p^{\mu}} & m\\
                \end{array} \right ] = 
		\slashed{p} + m,
\end{equation}
where, $\slashed{p} = h_{\mu\nu}\gamma^{\rho} e_{\rho}^{\nu}p^{\mu}$.
The results of quantum field theory can be reused with the generalization of the "dot" and the "Feynman slash" notations.

\subsubsection{Invariance of product of $\gamma$ matrices on $h_{00}$ }
\label{sec:innerprod}
If the polarization vector is normalized as 
\begin{equation}
	\label{eq:eps_norm}
	\epsilon_{\mu r} \epsilon^{\mu}_s = \delta_{rs}, 
\end{equation}
	where $r,s = 1,2$. Then,
\begin{eqnarray}
	h_{00}{\epsilon^0_r}^2 - \sum_{i=1}^3{\epsilon^i}^2 = 1,
\end{eqnarray}
or
\begin{equation}
	\sqrt{h_{00}}\epsilon^0_r = \sqrt{1 + \sum_{i=1}^3 {\epsilon^i}^2} .
\end{equation}
Similarly, if $k_{\mu}k^{\mu} = m^2$,
\begin{equation}
	\sqrt{h_{00}}k^0 = \sqrt{m^2 + \sum_{i=1}^3 {k^i}^2 }.
\end{equation}
Finally,
\begin{equation}
	\label{eq:spinor_innerprod}
	\epsilon_{\mu r}k^{\mu} = \sqrt{1+\sum_{i=1}^3 {\epsilon^i}^2} \sqrt{m^2 + \sum_{i=1}^3 {k^i}^2} - \sum_{i=1}^3 \epsilon^i_r k^i,
\end{equation}
which is independent of $h_{00}$.
This generally true for any inner product,
\begin{equation}
	\label{eq:spinor_innerprod_gen}
	a_{\mu }b^{\mu} = \sqrt{\alpha^2+\sum_{i=1}^3 {a^i}^2} \sqrt{\beta^2 + \sum_{i=1}^3 {b^i}^2} - \sum_{i=1}^3 a^i b^i,
\end{equation}
as long as it satisfies the constraints:
\begin{eqnarray}
	\label{eq:norm}
	a_{\mu} a^{\mu} = \alpha^2; ~~~~~
	b_{\mu} b^{\mu} = \beta^2,
\end{eqnarray}
where $\alpha$ and $\beta$ are constants.
Constraints like those mentioned above would be satisfied by the momenta of on-shell particles.
Finally, for terms like $\slashed{p} = \gamma^ae^{\mu}_a p^{\nu} h_{\mu\nu}$, we must square $M$ and obtain the trace. For odd number of $\gamma$ matrices, the trace is zero. 
For a product of two $\gamma$ matrix,


\begin{eqnarray}
	\label{eq:p1p2_1}
\nonumber	tr{\slashed{p}_1\slashed{p}_2} = tr\Big [ (\gamma^ae^{\mu}_a p_1^{\nu}h_{\mu\nu})  (\gamma^b e^{\alpha}_b p_2^{\beta}h_{\alpha\beta} ) \Big ] \\
\nonumber	=tr[\gamma^a\gamma^b] e^{\mu}_a e^{\alpha}_b p_1^{\nu} p_2^{\beta} h_{\mu\nu} h_{\alpha\beta}\\
\nonumber	= 4 \eta^{ab} e^{\mu}_a e^{\alpha}_b h_{\mu\nu}h_{\alpha\beta} p_1^{\nu}p_2^{\beta}\\
\nonumber	= 4 h^{\mu\alpha} h_{\mu\nu} h_{\alpha\beta} p_1^{\nu} p_2^{\beta} \\
	= 4 h_{\nu\beta}p_1^{\nu} p_2^{\beta}.
\end{eqnarray}
Finally,	
\begin{multline}
	\label{eq:p1p2_2}
	tr{\slashed{p}_1\slashed{p}_2} = 4p_{1\mu}p_2^{\mu} =
	4\Big [ \sqrt{m_1m_2 + \sum_{i=1}^3 {p_1^i}^2 } \\
	\times \sqrt{m_1m_2 + \sum_{i=1}^3 {p_2^i}^2} - \sum_{i=1}^3 p^i_1p^i_2 \Big ].
\end{multline}
Thus, from Eq.~\ref{eq:p1p2_2}, $tr{\slashed{p}_1\slashed{p}_2}$ is seen to be independent of $h_{00}$. 

For a product of $n$ $\gamma$ matrices,
\begin{multline}
	tr(\slashed{a}_1
	\slashed{a}_2 ...
	\slashed{a}_n ) = 
	(a_1.a_2) tr(\slashed{a}_3
	...
	\slashed{a}_n )\\
	-(a_1.a_3) tr(\slashed{a}_2
	\slashed{a}_4
	...
	\slashed{a}_n )
	... +(a_1.a_n) tr(\slashed{a}_2
	...
	\slashed{a}_{n-1} ).
\end{multline}
If $a_1$, $a_2$, ... , $a_n$ satisfy Eq.~\ref{eq:norm}, then,
based on Eq.~\ref{eq:spinor_innerprod_gen}, the inner product $(a_i.a_j)$ is independent of $h_{00}$. If the trace of $n-2$ gamma matrices on the R.H.S. is independent of $h_{00}$, then the trace of $n$ gamma matrices on the L.H.S. is independent of $h_{00}$.
But from Eq.~\ref{eq:p1p2_2}, we know that the trace of two gamma matrices is independent of $h_{00}$. Therefore, by mathematical induction, the trace of product of $n$ gamma matrices is independent of $h_{00}$ for even $n$.
For odd $n$, the trace of odd number of gamma matrices is zero.
Thus, the trace of product of any number of gamma matrices (even or odd) is independent of $h_{00}$.

For spin polarized spinors, the probability amplitude calculation can involve terms that include $\gamma^5 = \gamma^0\gamma^1\gamma^2\gamma^3$. Let us say we have a term like
\begin{math}
	\gamma^5\slashed{p}_1 ... \slashed{p}_n.
\end{math}
If $p_1$, ... $p_n$ are on shell momenta, then, the constraint in Eq.~\ref{eq:norm} is satisfied as $p_{r\,\mu}p_r^{\mu} = m^2$.
Subsequently, for the metric in Eq.~\ref{eq:g00_metric}, one has,
\begin{equation}
	\sqrt{h_{00}}p_r^0 = \sqrt{m^2 + \sum_{i=1}^3 p_r^i p_r^i}.
\end{equation}
Any term, $\slashed{p}_r$, is of the form $\gamma^be_b^{\mu}p_{r\,\mu} = \sqrt{h_{00}}\gamma^0 p_r^0 + \sum_i \gamma^ip^i_r$.
One may replace all occurrences of $\sqrt{h_{00}}p_r^0$ with $\sqrt{m^2 + \sum_i p_r^i p_r^i}$ in the expression $\gamma^5\slashed{p}_1 ... \slashed{p}_n$.
	Consequently, $\gamma^5\slashed{p}_1 ... \slashed{p}_n$ becomes independent of $h_{00}$.

\subsection {QED}
\label{sec:qed_interaction}
\begin{figure}
\includegraphics[width = 80mm,height = 80mm]{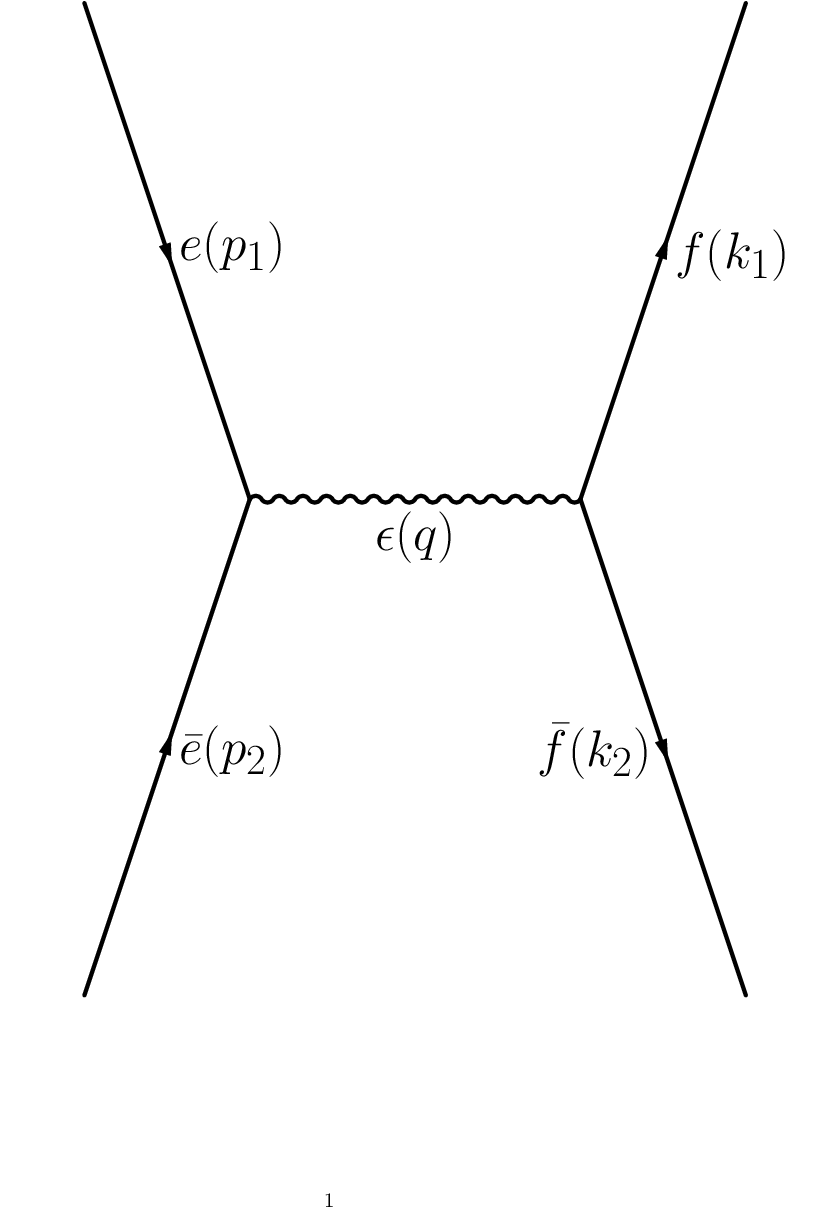}
	\caption{ $e^- + e^+ \rightarrow f + \bar{f}$}
\label{fig:mbg1}
\end{figure}

Let us consider the diagram of $e^- + e^+ \rightarrow e^- + e^+$ in Fig.~\ref{fig:mbg1} as a typical example. The results are then generalized for a generic QED process. We now evaluate the dependence of the S-matrix on the metric for the above diagram. 

Let us consider the invariant integral $\int \frac{d^4p}{(2\pi)^4} \delta(p_{\mu}p^{\mu} - m^2)$.
For the metric in Eq.~\ref{eq:g00_metric}, the invariant integral evaluates to: 
\begin{equation}
	\label{eq:invariant_int}
	\int \frac{d^4p}{(2\pi)^4} 2\pi \delta(p_{\mu}p^{\mu} - m^2) = \int \frac{d^3p}{(2\pi)^3} \frac{1}{2h_{00} p^0}.
\end{equation}
The normalization condition,
\begin{math}
	\langle p| q \rangle = (2\pi)^3 h_{00}p^0 \delta^3({\bf p}-{\bf q}),
\end{math}
gives:
\begin{equation}
	|p \rangle = \sqrt{h_{00}p^0} a^{\dagger}|0\rangle.
\end{equation}

Now, let us evaluate the term $\psi_I(x)|\bf{p},s\rangle$ for the metric in Eq.~\ref{eq:g00_metric} using the invariant integral in Eq.~\ref{eq:invariant_int},
\begin{multline}
	\label{eq:psi1}
	\psi_I(x)|{\bf p},s\rangle = 
	\int \frac{d^3p'}{(2\pi)^3} 
	\frac{1}{\sqrt{2h_{00}p'^{0}}} \\ 
	\times \Sigma_{s'} a^{s'}_{\bf{p'}} u^{s'}(p') \exp\left (-ip'_{\mu}x^{\mu}\right ) 
	\sqrt{2h_{00}p^{0}}a^{s\dagger}_{\bf{p}}|0\rangle.
\end{multline}
Using the anti-commutation relation, $\left \{a_{\bf{p'}},a_{\bf {p}}^{\dagger} \right \} = (2\pi)^3\delta^3(\bf{p'} - \bf{p}) \delta(s' - s)$, Eq.~\ref{eq:psi1} evaluates to 
\begin{equation}
\label{eq:psi2}
	\psi_I(x)|{\bf p},s\rangle = \exp(-ip_{\mu}x^{\mu}) u^s(p)|0\rangle.
\end{equation}

Following the standard process, the fermion propagator for the metric in Eq.~\ref{eq:g00_metric} can be seen to be
\begin{equation}
	\frac{i}{\slashed{p} - m} = \frac{i}{h_{\mu\nu} \gamma^r e^{\mu}_{r} p^{\nu} - m } = \frac{i}{\sqrt{h_{00}} \gamma^0 p^0 - \sum_j \gamma^j p^j - m}.
\end{equation}

We now consider the S-matrix. Let $S = I + iT$, where $T$ is the transition matrix. Then, 
\begin{multline}
	T(p_1,p_2 \rightarrow k_1,k_2) =\\
	\int \frac{d^4q}{(2\pi)^4} \int d^4x \int d^4y M (p_1,p_2 \rightarrow k_1,k_2)\\
\times	e^{-i(p_{1\mu} + p_{2\mu} - q_{\mu})x^{\mu}} e^{-i(-k_{1\nu} - k_{2\nu} + q_{\nu})y^{\nu}},
\end{multline}
where,
\begin{equation}
	M = \lambda^2 \overline{v}^{s'}(p_2)\gamma^a u^s(p_1) \frac{-i\eta_{ab}}{h_{\alpha\beta}q^{\alpha}q^{\beta}} \overline{u}^r(k_1) \gamma^bv^{r'}(k_2).
\end{equation}
Evaluating the integrals $\int d^4x$ and $\int d^4y$,
\begin{multline}
	\label{eq:qed1}
	T(p_1,p_2 \rightarrow k_1, k_2) = \int \Big [ \frac{d^4q}{(2\pi)^4} M (p_1,p_2 \rightarrow k_1,k_2)\\
\times	(2\pi)^4 \delta^4{(p_{1\mu} + p_{2\mu} - q_{\mu})} (2\pi)^4 \delta^4(-k_{1\mu} - k_{2\mu} + q_{\mu}) \Big ].
\end{multline}
Evaluating the $q$ integral, using the metric in Eq.~\ref{eq:g00_metric}, and noting that $\int d^4q = \int d^3q\int dq^0$, we obtain, 
\begin{multline}
	\label{eq:qed2}
	T = \frac{1}{h_{00}} M (p_1,p_2 \rightarrow k_1,k_2)\\
	\times (2\pi)^4 \delta^4{(p_{1\mu} + p_{2\mu} - k_{1\mu} - k_{2\mu})}.
\end{multline}
We see that integrating over a delta function brings down a factor of $\frac{1}{h_{00}}$ to the denominator.
To compute the final cross section, we need to calculate $M^2$.
\begin{multline}
	\label{eq:qed3}
	M^2 =  \frac{\lambda^4}{(q_{\alpha} q^{\alpha})^2} 
	tr\left ( \Sigma_{s s'} v^{s'}(p_2)\overline{v}^{s'}(p_2) \gamma^a u^s(p_1)\overline{u}^s(p_1)\gamma^c \right )\\
\times	tr\left ( \Sigma_{r r' } u^{r}(k_1)\overline{u}^{r}(k_1) \gamma_a v^{r'}(k_2)v^r(k_2)\gamma_c \right ),
\end{multline}
where $q = p_1 + p_2 = k_1 + k_2$.
For QED, $\lambda = e$, the charge of the electron. This becomes,
\begin{multline}
	\label{eq:qed4}
	M^2 =  \frac{\lambda^4}{(q_{\alpha} q^{\alpha})^2} 
	tr\left \{(\slashed{p_1} - m)\gamma^{\rho}(\slashed{p_2} + m) \gamma^j \right \}\\
	\times tr\left \{(\slashed{k_1} - m) \gamma_{\rho} (\slashed{k_2} + m) \gamma_j \right \},
\end{multline}
where, $\slashed{p} = h_{\mu\nu}\gamma^{\rho} e_{\rho}^{\nu}p^{\mu}$.
Since the trace of odd number of $\gamma$ matrices is zero, the above expression becomes:
\begin{multline}
	\label{eq:qed5}
	M^2 =  32\frac{\lambda^4}{(q_{\alpha} q^{\alpha})^2} 
	\Bigg [ (p_{1\mu}p_2^{\mu})(k_{1\nu}k_2^{\nu}) + (p_{1\mu}k_1^{\mu})(p_{2\nu}k_2^{\nu})\\
	+ (p_{1\mu}k_2^{\mu})(p_{2\nu}k_1^{\nu}) - m^2\Big \{ p_{1\mu}k_1^{\mu} + p_{2\mu} k_2^{\mu} \Big \} + 2m^4 \Bigg ].
\end{multline}
For a particle on the mass shell,
\begin{equation}
	\label{eq:qed_mass_shell}
	p_{\mu} p^{\mu} = m^2.
\end{equation}

For the metric in Eq.~\ref{eq:g00_metric}, Eq.~\ref{eq:qed_mass_shell} gives
\begin{equation}
	\label{eq:g00p0}
	\sqrt{h_{00}}p^0 = \sqrt{{\bf p.p} + m^2 }.
\end{equation}
For any inner product $p_{\mu} k^{\mu}$, we can use Eq.~\ref{eq:g00p0}, to find:
\begin{equation}
	\label{eq:qed_inner_prod}
	p_{\mu}k^{\mu} =  \sqrt{{\bf p.p} + m_1^2}\sqrt{{\bf k.k} + {m_2}^2} + {\bf p.k} + m_1m_2,
\end{equation}
where $m_1$ and $m_2$ are the masses of the respective particles.
The term $p_{\mu}k^{\mu}$ in Eq.~\ref{eq:qed_inner_prod} is seen to be independent of $h_{00}$. 
Thus, all terms in Eq.~\ref{eq:qed5}, like $q_{\alpha}q^{\alpha} = (p_{1\alpha} + p_{2\alpha})(p_1^{\alpha} + p_2^{\alpha})$, $p_{1\mu}p_2^{\mu}$, etc., are independent of $h_{00}$. 
Consequently, $M^2$ in Eq.~\ref{eq:qed5} is independent of $h_{00}$.
Therefore, we consider $M$ to be effectively independent of $h_{00}$.

In Eq.~\ref{eq:qed2}, we see that for 2 vertices, a factor of $\frac{1}{h_{00}}$ is applied. For every vertex added to Fig.~\ref{fig:mbg1}, we will have one more propagator, leading to an additional factor of $\frac{1}{h_{00}}$. There would also be one more additional factor of the coupling constant, $\lambda$.
Thus, if there are no loops, then for $v$ vertices, one may generalize:
\begin{equation}
	\label{eq:qed_vertex_scaling}
	\lambda^v \rightarrow \frac{\lambda^v}{h_{00}^{v-1}}.
\end{equation}

\subsection {Loops in Feynman diagram}
\label{sec:loops}
We now consider Feynman diagrams with loops.
\begin{figure}
\includegraphics[width = 80mm,height = 80mm]{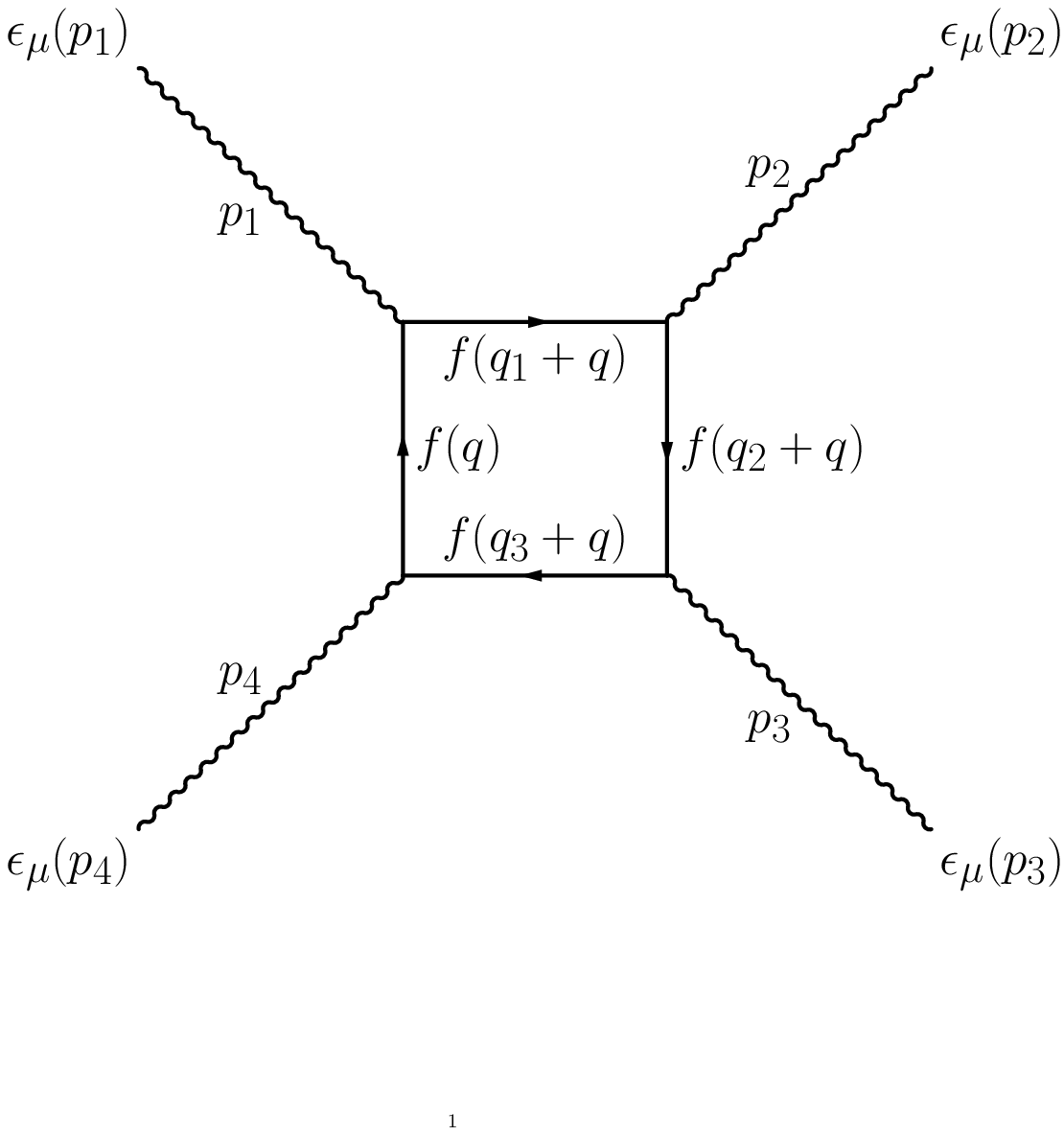}
	\caption{ A QED diagram with one loop.}
\label{fig:mbg_loop}
\end{figure}
Let us evaluate the loop in Fig.~\ref{fig:mbg_loop}.
\begin{multline}
	\label{eq:loop1}
	T(p_1,p_2, p_3, p_4) =\\
	\lambda^4 \int \frac{d^4q_1}{(2\pi)^4}\int \frac{d^4q_2}{(2\pi)^4}\int \frac{d^4q_3}{(2\pi)^4}\int \frac{d^4q}{(2\pi)^4} \\
	\times \epsilon(p_1) \frac{1}{\slashed{q}_1 + \slashed{q} - m}
	\epsilon(p_2) \frac{1}{\slashed{q}_2 + \slashed{q} - m} \\
	\times \epsilon(p_3) \frac{1}{\slashed{q}_3 + \slashed{q} - m}\epsilon(p_4) \frac{1}{\slashed{q} - m}\\
	\times (2\pi)^4 \delta^3( {\bf p_1} + {\bf q} - {\bf q_1} - {\bf q}) \delta(h_{00} (p_1^0 + q^0 - q_1^0 - q^0)) \\
	\times (2\pi)^4 \delta^3( {\bf p_2} + {\bf q_1} + {\bf q} - {\bf q_2} - {\bf q}) \delta(h_{00} (p_2^0 + q_1^0 + q^0 - q_2^0 - q^0)) \\
	\times (2\pi)^4 \delta^3( {\bf p_3} + {\bf q_2} + {\bf q} - {\bf q_3} - {\bf q}) \delta(h_{00} (p_3^0 + q_2^0 + q^0 - q_3^0 - q^0)) \\
	\times (2\pi)^4 \delta^3( {\bf p_4} + {\bf q_3} + {\bf q} - {\bf q}) \delta(h_{00} (p_4^0 + q_3^0 + q^0 - q^0)).
\end{multline}

This becomes,
\begin{multline}
	\label{eq:loop2}
	T(p_1,p_2,p_3,p_4) =\\
	\frac{\lambda^4}{(h_{00})^3 } 
	M(p_1,p_2,p_3,p_4) \delta^3({\bf p_4} + {\bf p_3} + {\bf p_2} + {\bf p_1} ) \\
	\times \delta\left \{ h_{00}(p_4^0 + p_3^0 + p_2^0 + p_1^0) \right \},
\end{multline}
and, 
\begin{multline}
	\label{eq:loop4}
	M(p_1,p_2,p_3,p_4) =\\
	tr \Big [ \int \frac{d^4 q }{(2\pi)^4}
	\epsilon(p_1) \frac{1}{\slashed{p}_1 + \slashed{q} - m}
	\epsilon(p_2) \frac{1}{\slashed{p}_1 + \slashed{p}_2 + \slashed{q} - m} \\
	\times \epsilon(p_3) \frac{1}{\slashed{p}_1 + \slashed{p}_2 + \slashed{p}_3 + \slashed{q} - m}
	\epsilon(p_4) \frac{1}{\slashed{q} - m} \Big ],
\end{multline}
where,
$d^4q = d^3q dq^0$.
Thus, so far, we have 
\begin{equation}
	\label{eq:loop5}
		\lambda^4 \rightarrow \frac{\lambda^4}{h_{00}^{3}}.
\end{equation}
	Now, if there were $v$ vertices in the loop in Fig.~\ref{fig:mbg_loop}, then, Eq.~\ref{eq:loop5} would have read 
\begin{equation}
\label{eq:loopscaling}
		\lambda^v \rightarrow \frac{\lambda^v}{h_{00}^{v-1}}.
\end{equation}
Now, $\slashed{q} = h_{\mu\nu} q^{\mu} e^{\nu}_a\gamma^a = \sqrt{h_{00}}q^0\gamma^0 + \sum_i q^i\gamma^i $.
Substituting $Q^0 = \sqrt{h_{00}}q^0$, the $d^4q$ integral becomes,
\begin{multline}
	\label{eq:loop6}
	M(p_1,p_2,p_3,p_4) =  \frac{1}{\sqrt{h_{00}}} tr \Big [ \int \frac{dQ^0}{2\pi}\\
	\times \int \frac{d^3q}{(2\pi)^3} 
	\epsilon(p_1) \frac{1}{\slashed{p}_1 + \slashed{Q} - m} 
	\epsilon(p_2) \frac{1}{\slashed{p}_1 + \slashed{p}_2 + \slashed{Q} - m}\\
	\times \epsilon(p_3) \frac{1}{\slashed{p}_1 + \slashed{p}_2 + \slashed{p}_3 + \slashed{Q} - m}
	\epsilon(p_4) \frac{1}{\slashed{Q} - m} \Big ],
\end{multline}
where,
$\slashed{Q} = Q^0\gamma^0 + \sum_iq^i\gamma^i$. 
Any occurrence of $\sqrt{h_{00}}p_r^0$ can be replaced by $\sqrt{h_{00}}p_r^0 = \sqrt{\sum_i p_r^i p_r^i + m^2}$, which is independent of $h_{00}$.
Thus, succinctly,
\begin{equation}
	\label{eq:loop7}
	M(p_1,p_2,p_3,p_4) =  \frac{M_0(p_1,p_2,p_3,p_4)}{\sqrt{h_{00}}},
\end{equation}
where $M_0$ is the value of $M$ when $h_{\mu\nu} = \eta_{\mu\nu}$.
Thus, we see that a loop introduces a factor $\frac{1}{\sqrt{h_{00}}}$.
If we had a boson propagator on one of the edges instead of a fermion propagator in the loop, then also the factor remains as $\frac {1}{\sqrt{h_{00}}}$.
For $l$ loops, there will $l$ $\frac{d^4q}{(2\pi)^4}$ integrals, which will result in a scale factor of $\frac{1}{(\sqrt{h_{00}})^l}$, i.e.,.
\begin{equation}
	\label{eq:loop8}
	M = \frac{M_0}{(\sqrt{h_{00}})^l}.
\end{equation}
From Eqs.~\ref{eq:loopscaling} and~\ref{eq:loop8}
\begin{equation}
	\label{eq:loop9}
	T \rightarrow \frac{\lambda^v}{(h_{00})^{v + l/2 - 1}} T_0,
\end{equation}
where, $T_0$ is the value of $T$ when $h_{\mu\nu} = \eta_{\mu\nu}$.
We can attribute the scaling factor of $\frac{1}{h_{00}^{v + l/2 -1}}$ to a scaling in $\lambda$ instead of $T$.
Combining Eqs.~\ref{eq:qed_vertex_scaling}, \ref{eq:loopscaling} and \ref{eq:loop9}, we may say that for QED, $\lambda$ scales as:
\begin{equation}
	\label{eq:qed_scaling}
	\lambda^v \rightarrow \frac{\lambda^v}{(h_{00})^{v + l/2 - 1}}.
\end{equation}

\subsection{QCD}
\label{sec:qcd}
We now look at a typical QCD diagram, and then attempt to generalize the results for a generic QCD process.
\begin{figure}
     \begin{subfigure}[b]{0.79\columnwidth}
         \centering
\includegraphics[width = \columnwidth,height = 67mm]{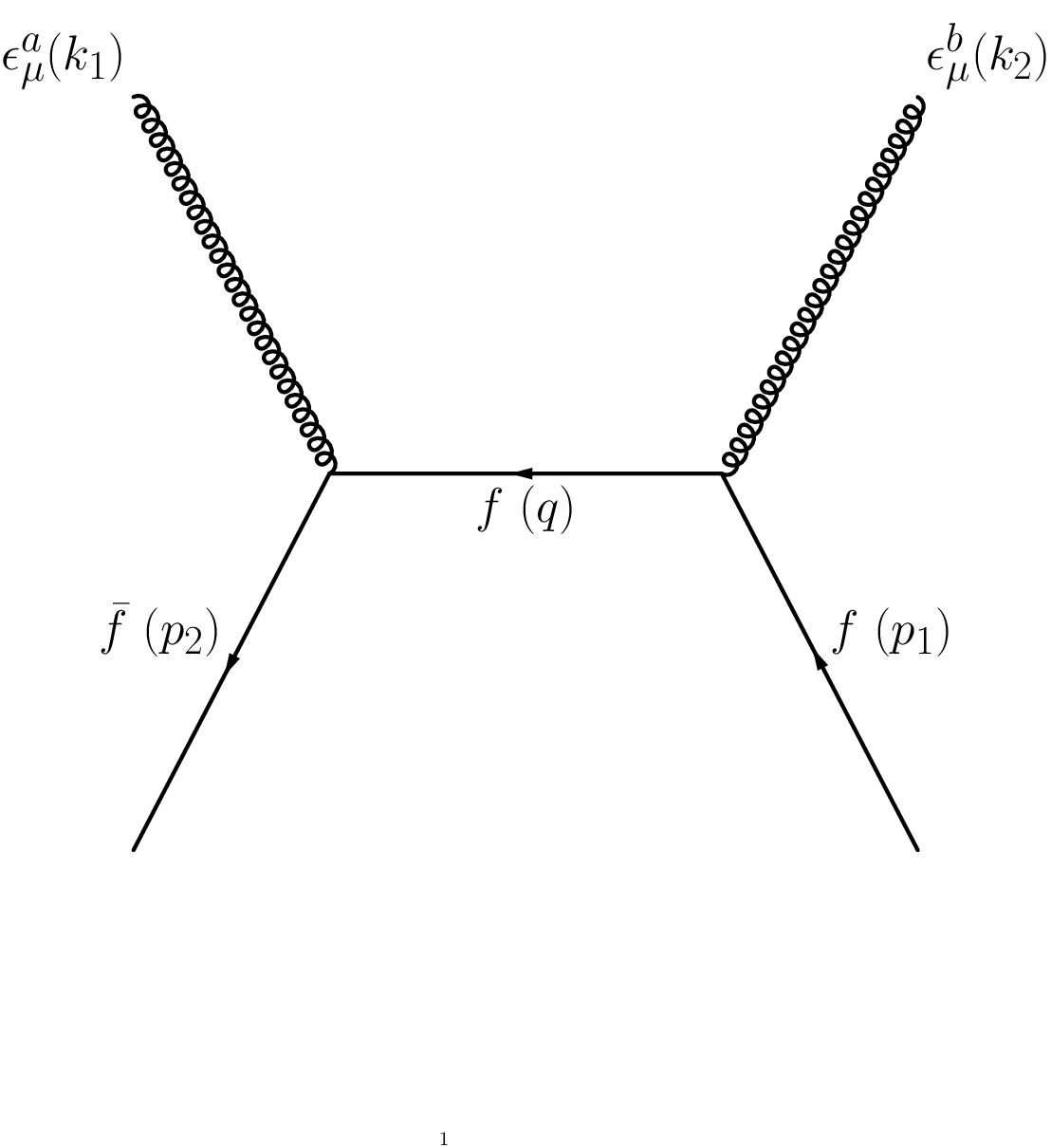}
	 \caption{ }
     \label{fig:mbg2}
     \end{subfigure}

     \hfill
     \begin{subfigure}[b]{0.79\columnwidth}
         \centering
\includegraphics[width = \columnwidth,height = 67mm]{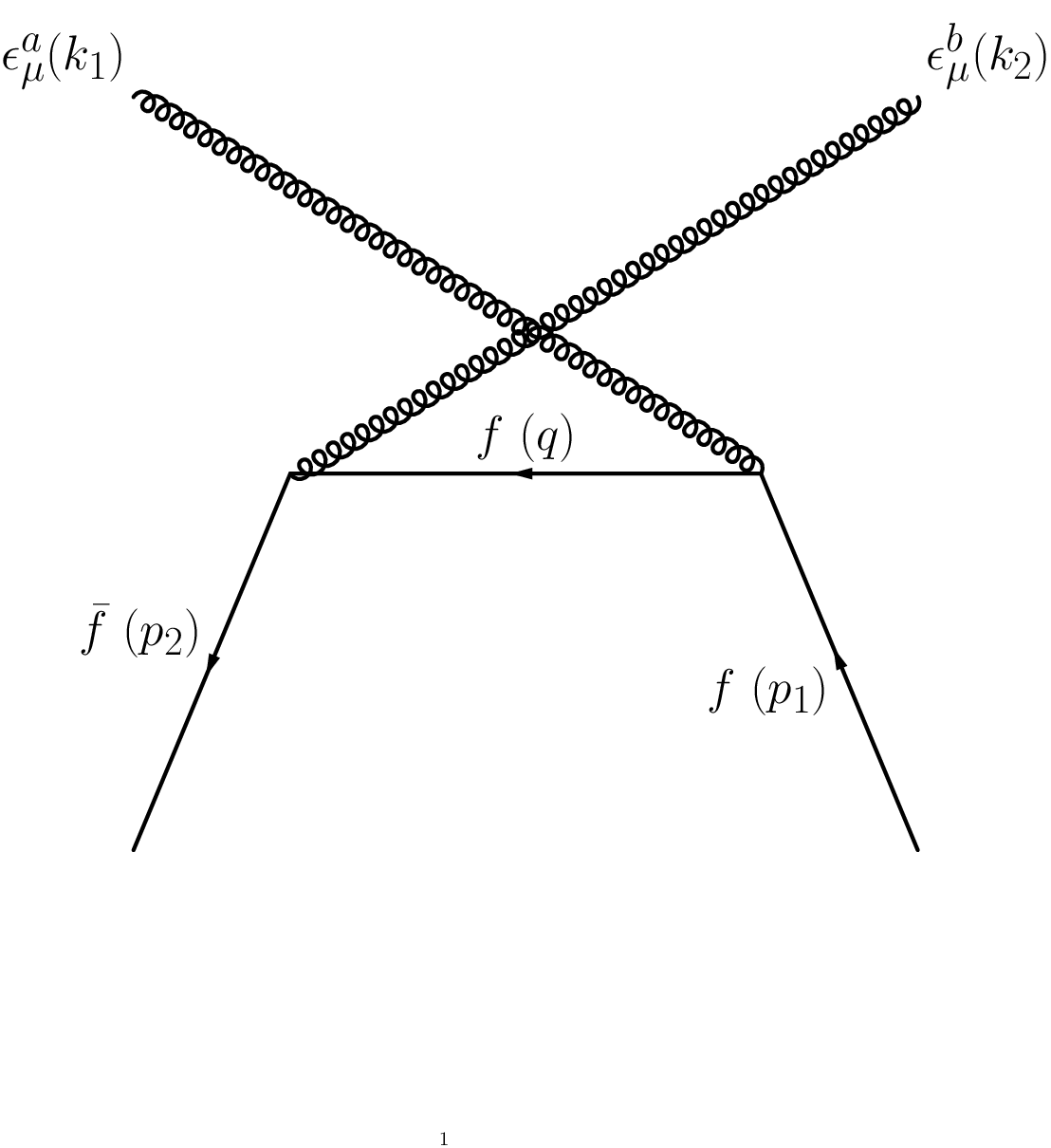}
	 \caption{ }
     \label{fig:mbg3}
     \end{subfigure}
     \hfill
     \begin{subfigure}[b]{0.79\columnwidth}
         \centering
\includegraphics[width = \columnwidth,height = 67mm]{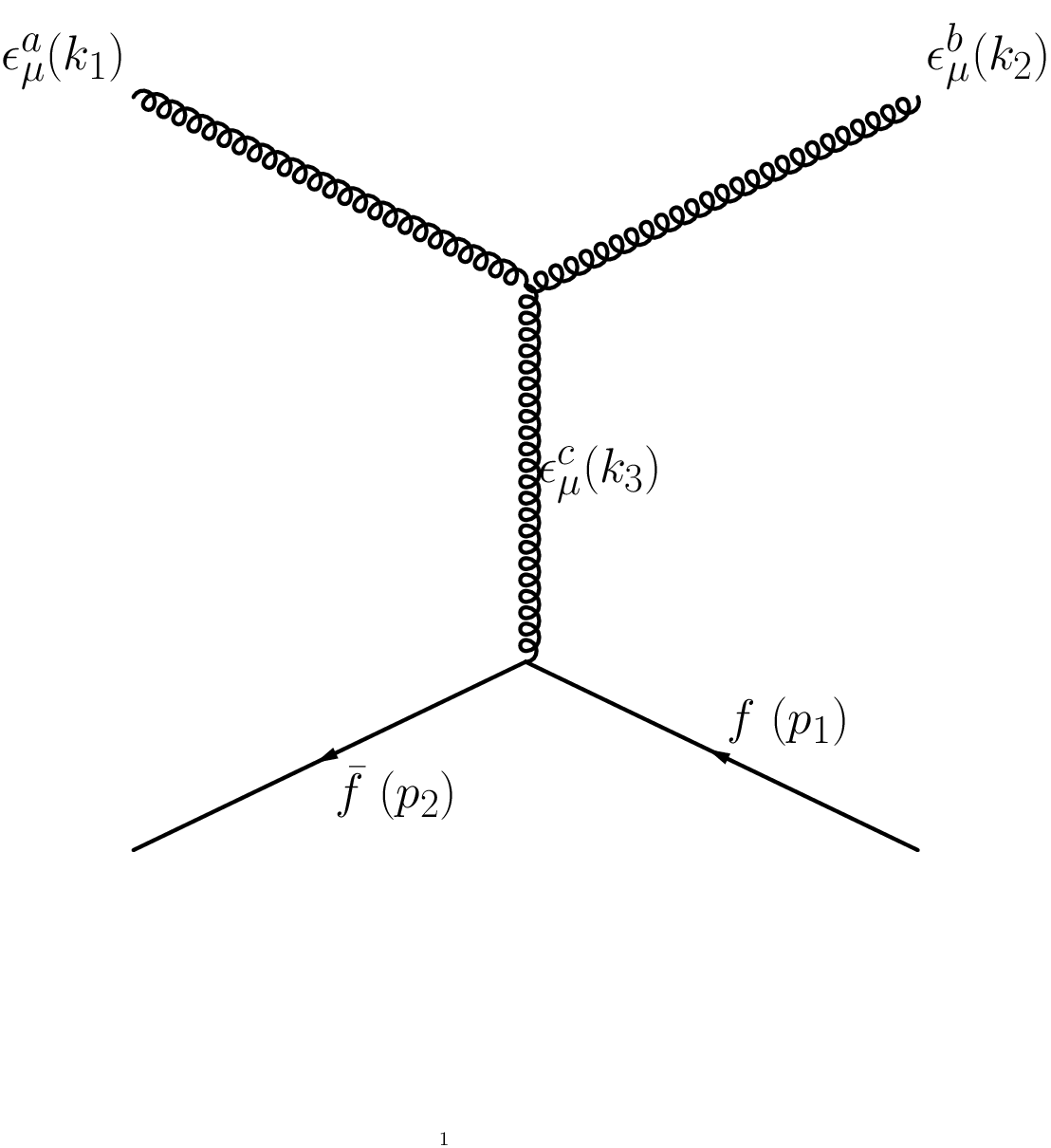}
	 \caption{ }
     \label{fig:mbg4}
     \end{subfigure}
     \caption{ $f^- + \bar{f}^+ \rightarrow g + g$}
\end{figure}

Sum of Figs.~\ref{fig:mbg2} and~\ref{fig:mbg3} gives:
\begin{multline}
	M_{12} = (ig)^2 \overline{v}(p_2) \Big \{ \gamma^{\mu}t^a \frac{i(\slashed{p_1} - \slashed{k_2} + m)}{(p_1 - k_2)^2 - m^2} \gamma^{\nu} t^b \\
	+ \gamma^{\nu}t^b \frac{i(\slashed{k_2} - \slashed{p_2} + m)} {(k_2 - p_2)^2 - m^2}\gamma^{\mu}t^a \Big \} u(p_1) \epsilon_{\mu}^*(k_1) \epsilon_{\nu}^*(k_2),
\end{multline}
where $\slashed{p} = h_{\mu\nu} p^{\mu}e^{\nu}_r \gamma^r$, with $r$ = spinor index, $e^{\nu}_r$ = vierbein,
and, $\gamma^{\mu} = \gamma^a e^{\mu}_a$.

From Fig.~\ref{fig:mbg4}, we have,
\begin{multline}
	M_3 = (ig) \overline{v}(p_2)\gamma_{\rho} t^c u(p_1) \frac{-i}{k_3^2} \epsilon_{1\,\mu}^{*}(k_1)\epsilon_{2\,\nu}^*(k_2)\\
	\times gf^{abc} \Big [ g^{\mu\nu}(k_2 - k_1)^{\rho} + g^{\nu\rho}(k_3 - k_2)^{\mu} + g^{\rho\mu}(k_1 - k_3)^{\nu} \Big ]. 
\end{multline}
\begin{multline}
	= ig^2f^{abc}t^c \overline{v}(p_2) \Bigg [ \Big \{ (\slashed{k}_2 - \slashed{k}_1 ) \epsilon_1^*(k_1).\epsilon_2^*(k_2) \Big \} \\
	+ \Big \{ \slashed{\epsilon}_2^*(k_2) \epsilon^*(k_1).(k_3 - k_2) \Big \} \\
	+ \Big \{ \slashed{\epsilon}_1^*(k_1) \epsilon_2^*(k_2).(k_1 - k_3) \Big \} \Bigg ] u(p_1) \frac{(-i)}{k_{3\mu} k^{\mu}_3},
\end{multline}
where, $\slashed{\epsilon}^*(k) = h_{\mu\nu}\epsilon^{\mu*}(k) e^{\nu}_r \gamma^r$.
and, $\slashed{k} = h_{\mu\nu}k^{\mu} e^{\nu}_r \gamma^r$.

Let $M^2 = (M_{12} + M_3)^2$. $M^2$ would consist of 
\begin{enumerate}
	\item sum of trace of products of $\gamma$ matrices, and involving 
		$p_1$, $p_2$, $k_1$,$k_2$, $\epsilon_1(k_1)$ and $\epsilon_2(k_2)$.
	\item sum of product of SU(3) generators $t^a$, $t^b$ and $f^{abc}$.
\end{enumerate}
The variables, $p_1$, $p_2$, $k_1$, $k_2$, are all on-shell momenta and hence would satisfy the constraints in Eq.~\ref{eq:norm}. 
$\epsilon_1(k_1)$ and $\epsilon_2(k_2)$ also meet the constraint in Eq~\ref{eq:norm} (see Eq.~\ref{eq:eps_norm}). 
Hence, as seen in Sec.~\ref{sec:innerprod}, the first item, i.e., the trace of the product of $\gamma$ matrices, is independent of $h_{00}$. The second one i.e., trace of the product of SU(3) group matrices $t^a$, $t^b$ etc is also independent of $h_{00}$, as these matrices are SU(3) generators and do not contain any component of the metric tensor.

Consequently, $M^2$ and therefore, effectively, $M$, does not lead to any additional $h_{00}$ factor for QCD. 
These arguments are valid for any standard QCD calculation.
Similar to Eq.~\ref{eq:qed2} in the QED case, scaling factors of $\frac{1}{h_{00}}$ would be introduced while integrating the delta functions.
Therefore, we can conclude that the scaling of $\lambda$ in the matrix, $T(p_1, p_2 \rightarrow k_1, k_2)$, would be the same as that of QED (Eq.~\ref{eq:qed_scaling}), i.e., 
\begin{equation}
	\label{eq:final_qcd_scaling}
	\lambda^v \rightarrow \frac{\lambda^v}{h_{00}^{v + l/2  - 1}}.
\end{equation}

\subsection{Electroweak theory }
\label{sec:weak}
For the current purpose, the calculations of the weak force are similar to those of QCD calculations.
The SU(3) generators $t^a$ are replaced by the SU(2) generators $\frac{1}{2} \sigma^a$, where the $\sigma^a$ matrices are the Pauli spin matrices.
	There are nuances related to the left-handed and right-handed fermions, which are not present in QED or QCD. 
	The left and the right-handed particles live in different representations of the gauge group.
	However, these do not impact our results on the scaling of $\lambda$.
	For our calculations, the arguments for the weak force are similar to those of QCD. We then end up with the scaling factor:
\begin{equation}
	\label{eq:final_weak_scaling}
	\lambda^v \rightarrow \frac{\lambda^v}{h_{00}^{v + l/2  - 1}}.
\end{equation}
\subsection{The Higgs Mechanism}
\label{sec:higgs}
Let $\phi$ be a scalar field, and let the covariant derivative of $\phi$ be~\citep{PnS}:
\begin{equation}
	\label{eq:higgs1}
	D_{\mu} \phi = \left ( \partial_{\mu} - i\lambda A^a_{\mu}\tau^a -i\frac{1}{2}\lambda'B_{\mu} \right ) \phi,
\end{equation}
where $A_{\mu}^a$ and $B_{\mu}$ are SU(2) and SU(1) gauge bosons.
$\lambda$ and $\lambda'$ are two different coupling constants.
Let $\phi$ be given as:
\begin{equation}
                \phi = \left ( \begin{array}{c }
			0 \\
			\sV \\
                \end{array} \right ),
\end{equation}
where, $\sV$ is the vacuum expectation value of the scalar field $\phi$.
Apart from the partial derivative terms, the Lagrangian arising from the kinetic term, $(D_{\mu}\phi)^{\dagger}D^{\mu} \phi$, would contain,
\begin{multline}
	\label{eq:higgs2}
	\Delta L = \frac{1}{2} (0~~~~\sV)
	\left ( \lambda \sum_a A_{\mu}^a\tau^a + \frac{1}{2} \lambda'B_{\mu} \right )\\
	\left ( \lambda \sum_a A^{\mu a} \tau^a + \frac{1}{2} \lambda' B^{\mu} \right ) 
                \left ( \begin{array}{c }
			0 \\
			\sV \\
                \end{array} \right ).
\end{multline}
We note that the index $a$ is a spinor index, and not a space-time index, and hence, it is not affected by the metric in Eq.~\ref{eq:g00_metric}.
Substituting $\tau^a = \frac{1}{2} \sigma^a$, we obtain
\begin{equation}
	\label{eq:higgs3}
	\Delta L = \frac{1}{2} \frac{\sV^2}{4} \left [ \lambda^2 A_{\mu}^1 A^{\mu 1} + \lambda^2 A_{\mu}^2 A^{\mu 2} + (-\lambda A_{\mu}^3 + \lambda'B_{\mu})^2 \right ].
\end{equation}

Let us redefine the fields,
\begin{eqnarray}
	\label{eq:higgs4}
	W_{\mu}^{\pm} = \frac{1}{\sqrt{2}} (A_{\mu}' \mp iA_{\mu}^2), \\
	Z_{\mu}^{0} = \frac{1}{\sqrt{\lambda^2 + \lambda'^2}} (\lambda A_{\mu}^3 - \lambda' B_{\mu}), \\
	A_{\mu} = \frac{1}{\sqrt{\lambda^2 + \lambda'^2}} (\lambda' A_{\mu}^3 + \lambda B_{\mu}).
\end{eqnarray}
The fields $W_{\mu}^{\pm}$ and $Z_{\mu}^0$ can be recognized as $W$ bosons and $Z$ bosons, respectively.
Then,
\begin{multline}
	\label{eq:higgs8}
	\Delta L = \frac{1}{2} \Bigg [ m_w^2 \Big \{ h_{\mu\nu}W^{\mu +*}W^{\nu +} + h_{\mu\nu}W^{\mu -*} W^{\nu -} \Big \} \\
	+ m_z^2\Big \{h_{\mu\nu} Z^{\mu 0*}Z^{\nu +} \Big \} \Bigg ],
\end{multline}
where, 
\begin{equation}
	\label{eq:wboson_mass}
m_w^2 = \frac{\sV^2}{4} \lambda^2 ,
\end{equation}
	and  
\begin{equation}
	\label{eq:zboson_mass}
	m_z^2 = \frac{\sV^2}{4} (\lambda^2 + \lambda'^2). 
\end{equation}
	The field, $A_{\mu}$, is massless.
	Very importantly, $m_w$ and $m_z$ are not impacted by $h_{00}$, and remain invariant to any tweaking of $h_{00}$.
In order to find the relation between $\lambda$ and time,
we substitute, $m_w = \sqrt{h_{00}} E_w$. 
\begin{equation}
	\label{eq:higgs9}
	\sqrt{h_{00}} E_w = \frac{1}{2} \lambda \sV \Rightarrow E_w = \frac{1}{2} \left (\frac{\lambda}{\sqrt{h_{00}}} \right ) \sV. 
\end{equation}
Similarly,
\begin{equation}
	\label{eq:higgs10}
	E_z = \frac{1}{2} \sqrt { \left (\frac{\lambda}{\sqrt{h_{00}}} \right )^2 + \left (\frac{\lambda'}{\sqrt{h_{00}}} \right )^2 } \sV. 
\end{equation}
Thus, scaling of $E_w$ and $E_z$ w.r.t. $h_{00}$ implies that the conjugate variable, time, also scales with $h_{00}$.

This leads to an equivalence between the scaling of time and the scaling of the coupling constant $\lambda$, although mass remains unaffected.
As a note, we have taken $c=1$ in all the above equations.

\subsection{Origin of mass}
\label{sec:origin}
We had discussed the Higgs mechanism for generating mass in Sec.~\ref{sec:higgs}. 
The Higgs mechanism is the source of mass for elementary particles, but not all masses.
For example, let us take the mass of a proton or a neutron, which makes up more or less the entire baryonic mass in this universe.
These masses have their origin in the QCD interactions. 
Essentially, they are the energy generated by the sea of virtual quarks and gluons.
Since these masses have their origin in the QCD interactions, the relation between the coupling constant $\lambda$ and time would already be covered by the QCD treatment in Sec.~\ref{sec:qcd}.
The mass however remains invariant with respect to $h_{00}$. To see why, let $t^0 \rightarrow \sqrt{h_{00}}t^0$. Then, the conjugate variable, $E^0 \rightarrow \frac{E^0}{\sqrt{h_{00}}}$.
This implies, $E_0 \rightarrow \sqrt{h_{00}}E^0$. Subsequently, the mass, $m = \sqrt{E^0E_0}$, which remains invariant.

\subsection{Final scaling and propagation of time}
\label{sec:finalscaling}
Based on Sec.~\ref{sec:spinor},~\ref{sec:qed_interaction},~\ref{sec:loops},~\ref{sec:qcd}, and~\ref{sec:weak}, we can finally arrive at the scaling of the coupling constant for QED, QCD and the Weak interactions:
\begin{equation}
	\label{eq:scaling1}
	\lambda^v \rightarrow \frac{\lambda^{v}}{h_{00}^{v+l/2-1}}.
\end{equation}

Let $M(h_{00})$ be the probability amplitude with the metric in Eq.~\ref{eq:g00_metric} and $M(1)$ be the probability amplitude with $h_{00}=1$.
The probability amplitude of any sum of diagrams, $M$, for QED, QCD and the weak force, can be written as:
\begin{equation}
	S_M(p_1,p_2) = \frac{M(h_{00})}{M(1)} = \frac{\sum_v\sum_l M_{vl}(p_1,p_2)\frac{\lambda^v}{h_{00}^{v + l/2 - 1}}} {\sum_v\sum_l M_{vl}(p_1,p_2)\lambda^v},
\end{equation}
where, $M_{vl}(p_1,p_2)$ is the probability amplitude for a diagram with $v$ vertices and $l$ loops. 

\begin{figure}
\includegraphics[width = 80mm,height = 80mm]{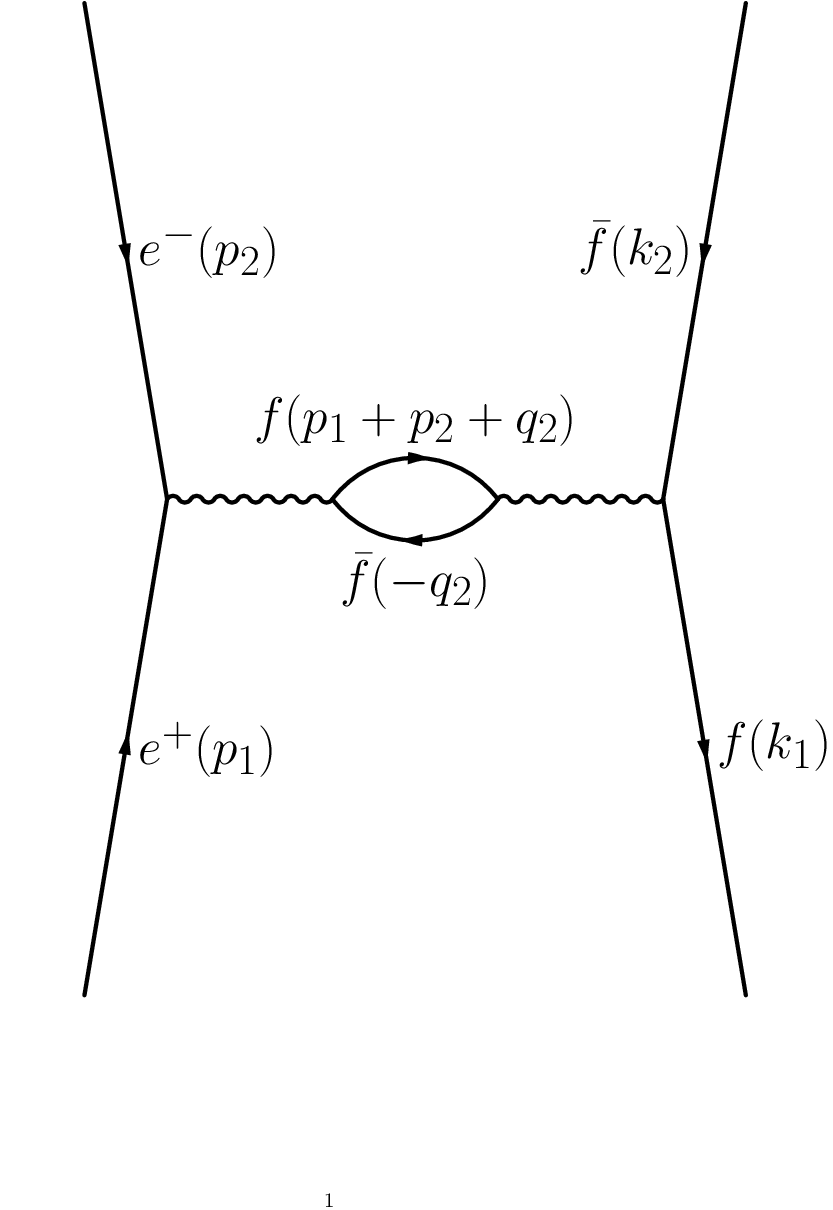}
	\caption{ A typical NLO diagram for $e^- + e^+ \rightarrow f + \bar{f}$}
\label{fig:mbg6_nlo}
\end{figure}
To understand further, let us look at the Next to Leading Order (NLO) diagram in Fig.~\ref{fig:mbg6_nlo}, which would give us $M_{41}(p_1,p_2)$ and compare it with $M_{20}(p_1,p_2)$ in diagram Fig.~\ref{fig:mbg1}.
The additional loop in the diagram in Fig.~\ref{fig:mbg6_nlo}, gives the factor,
\begin{equation}
	\lambda^2 \int \frac{d^4q_2}{(2\pi)^4} \frac{1}{(\slashed{p}_1 + \slashed{p}_2 + \slashed{q}_2 + m_f) (-\slashed{q}_{2}) }.
\end{equation}
As a result, $M_{41}(p_1,p_2)$ and ${M_{20}(p_1,p_2)}$ are different functions of $p_1$ and $p_2$. Consequently, $S_M(p_1,p_2)$ will vary with $p_1$ and $p_2$.

Thus, as time scales with $\sqrt{h_{00}}$, $S_M(p_1,p_2)$ scales in a non-trivial way. This is shown to be true even in the simple case where $h_{00}$ is (almost) a constant in a local region.
What does this imply? Let us say that we have two systems, $S_1$ and $S_2$, which have different particle compositions in terms of $p_1$ and $p_2$. 
Thus, even though the coupling constant $\lambda$ is scaled by the same constant factor in both cases, time can still evolve in a different way.

\subsection{Interpretation and relation of interaction to MBG}
\label{sec:interpretation}
So far, we have examined the relation between the coupling constant, $\lambda$ and time, $t$.
  The entropy of a system is dependent on the interactions and state variables such as momentum and position.
  The coupling constant, $\lambda$, modifies the interaction. 
Therefore, when there is a relation between coupling constants and time, there exists a relation between entropy and time.

From Heisenberg's uncertainty relation, $\Delta E \Delta t \ge \frac{h}{2\pi}$.
If $\Delta t$ increases due to time dilation, then $\Delta E$ decreases. 
This implies that the thermal energy (interaction energy), $\langle E \rangle$, and temperature decrease. Consequently, the inverse temperature, $\beta$ increases.  
This leads to a different inverse temperature, $\beta$, in regions I, II and III, in Fig.~\ref{fig:mbg_regions}.
We then infer a direct relation between the coupling constant, $\lambda$, time, and the inverse temperature, $\beta$.

The MBG theory incorporates the variation of $\beta$ in its laws of gravity.
The inverse temperature, $\beta({\bf x},t)$, of a system is a manifestation of the interactions between the constituents of the system.
As seen in Eq.~\ref{eq:define_s}, $\beta({\bf x},t) = s({\bf x},t)\beta_c$. We capture all the variations in $\beta({\bf x},t)$ in $s({\bf x},t)$.
The terms $\frac{k}{s}\nabla_{\nu}\nabla_{\nu}s$ and $\frac{k}{s}\nabla^{\alpha}\nabla_{\alpha} s$ capture the variation in the interactions (in the form of variation in the inverse temperature) within a system. Subsequently, Eq.~\ref{eq:gal3} captures the impact of variation in the interaction on space-time.

To wrap up, during interaction, the interacting wave-functions evolve. 
When the coupling constant is smaller, the interaction will evolve more slowly, which means time will be slower.
The interaction is not a point impact as in a classical picture. 
The wave-functions are smeared in space-time, and the interaction is a smeared continuous process.
All particles are interacting with each other.
The system of interacting wave-functions can be viewed as smeared coalesced wave-functions evolving together. These interacting wave-functions are spread in the fabric of space-time and manifests space-time itself.
It seems that interactions and time evolution are two sides of the same coin.

\section{Entropy and gravity}
\label{sec:single_particle_MBG}
\subsection{GR, MBG and entropy}
\label{sec:GR_MBG}
Let's now examine the original question of how a single particle or a collection of non-interacting particles in the universe affects space-time in the context of Einstein's gravity.
\begin{equation}
	\label{eq:GR}
	R_{\mu\nu} - \frac{1}{2}g_{\mu\nu}R = \frac{8\pi G}{c^4} T_{\mu\nu}.
\end{equation}
The metric $g_{\mu\nu}$ is a real physical metric, but $h_{\mu\nu}$ is mainly a mathematical knob to adjust interaction and determine the relationship between interaction and time.
For a single particle, or non-interacting particles, pressure, $P=0$. 
We take the four-velocity = $u^{\mu} = (1,0,0,0)$.
Then, 
\begin{equation}
	\label{eq:T00}
	T_{00} = \frac{8\pi G}{c^2} \rho, 
\end{equation}
and all other components of $T_{\mu\nu}$ are 0.
If we look the mass term in Eqs.~\ref{eq:wboson_mass} and~\ref{eq:zboson_mass}, mass is independent of $h_{00}$.
Mass remains unaffected by the scaling of time via $h_{00}$.
This leads to an inconsistency in Eq.~\ref{eq:GR}.
To understand why, let us first model a system of non-interacting particles.
Since the forces other than gravity are dominant over gravity, let us consider only the other three fundamental forces for interaction for now. We shall consider gravity separately. 
A single particle or a collection of non-interacting single particles can be modeled by taking the limit $h_{00} \rightarrow \infty$
If we take $h_{00} \rightarrow \infty$, then, from Eq.~\ref{eq:qed_scaling},
	and~\ref{eq:final_qcd_scaling}
$\frac{\lambda}{(h_{00})^{\frac{v + l/2 - 1}{v}} } \rightarrow 0$, i.e. all interactions vanish.
This limiting result is true as long as $v + l/2 -1 > 0$. Since the minimum value of $v=2$, the limiting result is true for all diagrams, regardless of the number of vertices or loops in a Feynman diagram.
Let us refer the metric,
\begin{equation}
	\label{eq:metric}
	ds^2 = h_{00}dt^2 - dx^2 - dy^2 - dz^2.
\end{equation}
In the limit, $h_{00} \rightarrow \infty$, for any finite $ds^2$, $dt \rightarrow 0$. 
Consequently, time becomes meaningless.
Hence, in the limit $h_{00} \rightarrow \infty$, the R.H.S. of Eq.~\ref{eq:GR} remains, well defined, but the time variable in the L.H.S. becomes ill-defined. 
One would expect the derivatives of $g_{00}$ to vanish or $g_{00}$ to become ill-defined, but such is not the case.
Thus, general relativity provides a meaningful space-time metric for a collection of non-interacting particles. 
for which time ceases to be meaningful. 
Even when the number of particles equals 1, when time becomes more ambiguous, general relativity continues to provide a well-defined time.

Let us now shift our attention to gravitational coupling. 
For the sake of simplicity, let gravity be the sole force of interaction and all other coupling constants be zero.
By taking the gravitational coupling, $G\rightarrow 0$, the gravitational interaction between the particles reduces to zero. This implies that the R.H.S. of Eq.~\ref{eq:GR} becomes zero. Consequently, the L.H.S., becomes zero. Time stops passing. The Einstein equation seems to capture the relationship between gravitational coupling and time. It fails to capture the relation between interactions due to the other fundamental forces and time.  
This is because gravity couples the particles to each other via the fabric of space-time, unlike other fundamental forces.
One may however argue that Einstein's equation fails to fully capture the effect of entropy due to gravitational interactions itself. 
For example, MBG explains galaxy rotation curves by also taking into account the variation in the gravitational interaction between stars~\citep{gans8, gans8E}.

Before venturing into the MBG theory, let's perform a thought experiment that doesn't ignore gravity while analyzing other forces.
Let us conceive a universe composed of only naked top quarks, with the electromagnetic coupling, $e$, reduced so that the repulsion between top quarks matches the gravitational attraction at least to a first order. 
Consequently, the top quarks interact through the strong force. 
A top quark soup is formed, and time flows in the soup. Entropy is non-zero.
Now, for the strong force, we take $h_{00}\rightarrow \infty$. Then the strong force coupling, $\lambda \rightarrow 
\frac{\lambda}{(h_{00})^{\frac{v + l/2 - 1}{v}} } \rightarrow 0$, i.e. all interactions due to strong force vanish.
As per preceding arguments, time vanishes, but the mass of the top quark does not. 
Intuitively, if all particles in the universe are stationary, time does not flow.
Equation~\ref{eq:GR}, however, continues to exhibit a well defined space-time. 
This again points to a discrepancy while using Einstein's gravity.
The problem lies in the absence of entropy in some form in Eq.~\ref{eq:GR}. 

Let us now explore this issue for the other three fundamental forces via the entropic/thermal term,
$\frac{k}{s}\nabla_{\mu}\nabla_{\nu}s - \frac{k}{s}g_{\mu\nu}\nabla^{\alpha}\nabla_{\alpha} s$,
in MBG.
Let us rearrange the terms in the MBG equation, Eq.~\ref{eq:gal3}, as:
\begin{equation}
	\label{eq:mbg_eq1}
	R_{\mu\nu} - \frac{1}{2}g_{\mu\nu}R^{(4)}  
	= \frac{8\pi G}{c^4} \rho u_{\mu}u_{\nu} 
+ \frac{k}{s}\nabla_{\mu}\nabla_{\nu}s - \frac{k}{s}g_{\mu\nu}\nabla^{\alpha}\nabla_{\alpha} s.
\end{equation}
		    Here, $k$ represents the degree or extent of equilibration. 
In essence, the 
$\left ( \frac{k}{s}\nabla_{\mu}\nabla_{\nu}s - \frac{k}{s}g_{\mu\nu}\nabla^{\alpha}\nabla_{\alpha} s \right )$ 
term is what captures interactions.
In order to analyze the behavior of L.H.S. of Eq.~\ref{eq:mbg_eq1}, due to 
$\frac{k}{s}\nabla_{\mu}\nabla_{\nu}s - \frac{k}{s}g_{\mu\nu}\nabla^{\alpha}\nabla_{\alpha} s$,
we temporarily remove the $\rho u_{\mu}u_{\nu}$ term in Eq.~\ref{eq:mbg_eq1}. 
Then, we obtain:
\begin{equation}
	\label{eq:mbg_eq2}
	R_{\mu\nu} - \frac{1}{2}g_{\mu\nu}R^{(4)} 
	= \frac{k}{s}\nabla_{\mu}\nabla_{\nu}s - \frac{k}{s}g_{\mu\nu}\nabla^{\alpha}\nabla_{\alpha} s.
\end{equation}

First let us consider the case I in Sec.~\ref{sec:intro}, i.e., the case of a single particle.
The factor, $k$ is the degree of thermalization. It models the amount of interaction between the particles~\citep{gans8,gans8E, gans9}. If there is only one particle, then there can be no interaction, and $k\rightarrow 0$.
In fact, in~\citep{gans8}, it was shown that $k\propto \sqrt{M_{galaxy}}$, in order for MBG to satisfy the Tully Fisher relation~\citep{tfr}.
Let us take $M_{galaxy} \approx N m$, where $N$ is the number of stars in the galaxy, and $m$ is the average mass of a star. Then, we obtain 
\begin{equation}
	\label{eq:tfr1}
	k\propto \sqrt{N}.
\end{equation}
The above relation is appropriate if $N$ is large.
However, for a one body system, there is obviously no interaction. Even for a two body system, it's a very deterministic motion between the two particles. There is no "variation" in the interaction, i.e., no random component. One needs at least a three body system to have a variation in interaction. Subsequently, a more appropriate form of Eq.~\ref{eq:tfr1} would be: 
\begin{equation}
	\label{eq:tfr2}
	k\propto \sqrt{N-2}~~~~  \forall N>=2.
\end{equation}
Since there are billions of stars in a galaxy, $N$ is significantly large.
For such large $N$, Eq.~\ref{eq:tfr2} is almost same as Eq.~\ref{eq:tfr1}. 

In Ref.~\cite{gans8}, it was argued that for wide binary stars, which is a two body system, MBG predicts a Newtonian gravity, i.e,, the entropic terms do not contribute. In fact, there are observational evidences, that WBS can in fact behave in a Newtonian manner~\citep{wbs3}.
If $k \rightarrow 0$, for a one or two body system,
then, $\frac{k}{s}\nabla_{\mu}\nabla_{\nu}s - \frac{k}{s}g_{\mu\nu}\nabla^{\alpha}\nabla_{\alpha} s$ vanishes.

We now consider the case II in Sec.~\ref{sec:intro}, i.e., the case of a collection of stationary non-interacting particles.
For a collection of non-interacting stationary particles, there is no temperature. i.e, its a vacuum. For zero temperature, there can be no thermal gradient, i.e., $\partial_{\mu} s \rightarrow 0$. 
Consequently, 
$\nabla_{\mu}\nabla_{\nu}s \rightarrow 0$ and $\nabla^{\alpha}\nabla_{\alpha} s \rightarrow 0$.
This argument, in the limit of the number of non-interacting stationary particles $\rightarrow 1$,  can also be applied to case I.

As a result of the above arguments, for either case I or case II, Eq.~\ref{eq:mbg_eq2} yields:
\begin{equation}
	\label{eq:GR0}
	R_{\mu\nu} - \frac{1}{2}g_{\mu\nu}R = 0.
\end{equation}
Thus, in the context of pure interactions, MBG does not modify the time for a single particle or a collection of stationary non-interacting particles.
However, MBG theory also contains the $T_{\mu\nu}$ term like Einstein's gravity, which in turn includes the mass density $\rho$. 
The mass term creates time ambiguity for MBG also. 

As a consequence of the issue discussed so far, at the beginning of time, if the universe is filled with massive scalar particles, then, taking $t \rightarrow 0$, would be hindered by a non-zero $\rho$, as long as $G \ne 0$.
This prompts us to consider a scalar field, $\hat{\phi}$, that is massless, with the scalar-scalar interaction driving the gravitational dynamics.
In the case of Einstein's gravity, the interaction is introduced in the form of an external potential $V(\hat{\phi})$~\citep{potential1, potential2}. This is the classical inflation model and we will not delve into this further.
In the case of the MBG theory, the interactions are captured by the entropic term, 
$\frac{k}{s}\nabla_{\mu}\nabla_{\nu}s - \frac{k}{s}g_{\mu\nu}\nabla^{\alpha}\nabla_{\alpha} s$, which gives the correct behavior in the case of zero interactions.
In the next section, Sec.~\ref{sec:friedmann}, we analyze a massless scalar in 5-D space-time-temperature. 
We also show that, if the entropic term, 
$\frac{k}{s}\nabla_{\mu}\nabla_{\nu}s - \frac{k}{s}g_{\mu\nu}\nabla^{\alpha}\nabla_{\alpha} s$, dominates, then the MBG theory predicts inflation as a possibility.
\subsection{Possible Boundary Cases}
\label{sec:counter_arguments}
Before we begin modeling the Friedman universe with MBG, let's explore some scenarios that may be boundary cases related to our current discussions.

We begin with the case of gravitons in a Bose-Einstein condensate, with a large occupancy number, $N \gg 1$, leading to the formation of a black hole~\cite{graviton}.
One might argue that a universe composed of only massless gravitons forming black holes implies the evolution of time.
Time evolves, as the state of the universe changes.
The question is whether this conflicts with the basic tenets of our discussion in any way.
As mentioned in Ref.~\cite{graviton}, black hole is considered as a bound state of gravitons. Any bound state implies interactions, and the keyword here is {\it interactions}, or non-zero coupling. 
As discussed in Sec.~\ref{sec:interaction}, we consider the limiting case of the coupling constant becoming zero and its impact on time.
The discussion in this paper doesn't conflict with a non-zero coupling constant that leads to time evolution.

Another case could be Heisenberg's uncertainty principle,
\begin{enumerate}
\item $\Delta p \Delta x \ge \frac{h}{4\pi}$,
and
\item $\Delta E \Delta t \ge \frac{h}{4\pi}$.
\end{enumerate}
Let us analyze whether the uncertainty principle negates the discussion related to Case II in Sec.~\ref{sec:intro}. If $\Delta p > 0$, then particles cease to be stationary, The question is would time flow in this case?
Also, would the uncertainty principle induce a non-zero entropy by virtue of different particles being in a different superposition of eigenstates?

We club the first case of $\Delta p \Delta x \ge \frac{h}{4\pi}$ with the second case, as momentum variation leads to energy variation. 
For the second case of $\Delta E \Delta t \ge \frac{h}{4\pi}$, let us consider the well known result of the Fourier transform of a Gaussian wave-packet,
\begin{multline}
\label{eq:fourier1}
	\frac{1}{\sqrt{2\pi}\sigma_e} \exp\left (- \frac{E^2}{2\sigma_e^2} \right ) \leftrightarrow  
	\exp \left (- \frac{t^2 \sigma_e^2}{2} \right ).
\end{multline}
The R.H.S. of Eq.~\ref{eq:fourier1} is a wave-packet of standard deviation $\sigma_t = \frac{1}{\sigma_e}$. We shall now see how $h_{00}$ impacts the above result.
\begin{multline}
	\int dt \frac{1}{\sqrt{2\pi}\sigma_t} \exp\left (- \frac{t^2}{2\sigma_t^2} \right ) \exp \left(-ih_{00}Et\right )\\
	= \int d\nu \frac{1}{\sqrt{2\pi}h_{00}\sigma_t} \exp\left (- \frac{\nu^2}{2h_{00}^2 \sigma_t^2} \right ) \exp \left(-iE\nu \right ),
\end{multline}
with $\nu = h_{00}t$.
This indicates that the new standard deviation = $\sigma_t' = h_{00}\sigma_t$. The standard deviation in the Fourier domain would be $\sigma_e' = \frac{\sigma_e}{h_{00}}$. For the no interaction limit, we take $h_{00} \rightarrow \infty$, which gives $\sigma_e' \rightarrow 0$ and $\sigma_t' \rightarrow \infty$. 
This is a restatement of an energy eigenstate remaining constant over infinite time, and in fact mirrors the scenario mentioned in case II in Sec.~\ref{sec:intro}.
A constant energy eigenstate implies there is no interaction, and the universe does not evolve over time.
While the time axis exists, the operational notion of time becomes ill-defined.
Thus, the discussions presented in case II in Sec.~\ref{sec:intro}, Sec.~\ref{sec:interaction} and Sec.~\ref{sec:GR_MBG} are consistent with the uncertainty principle.

\section{The Friedmann universe}
\label{sec:friedmann}
\subsection{The values of k at different era}
\label{sec:kvalues}
We first analyze a massless scalar, with a 5-D FLRW like metric.
Subsequently, we examine the MBG theory using the FLRW metric for the inflationary and deep matter eras in 
Sec.~\ref{sec:inflation_era} and Sec.~\ref{sec:matter_era} respectively.
As discussed in the previous section, 
we will show that the universe can be modeled as a pure interaction, via a massless scalar.

In a fully thermalized system, $k \sim -1$.  
We recollect from Ref.~\cite{gans8E}, that $k$ is negative, with the max absolute value of $|k| = 1$.
Let us elaborate a bit on $k \sim -1$. In the context of galaxy rotation curves in Ref.~\cite{gans8}, galaxy masses from $M_{galaxy}= 0.21\times 10^9 M_{\odot}$ to $M_{galaxy}= 67.75\times 10^9 M_{\odot}$ were analyzed, giving the range of $k = -2.4135\times 10^{-11} \sqrt{M_{galaxy}}$, from $-3.5\times 10^{-7}$ to $-6.3\times 10^{-6}$.
Hence, even a value of $k=-0.3$ can be deemed as close to -1!

For a fully thermalized system, the value of $|k|$ being maximum further augments the dominance of the entropic terms,
$\frac{k}{s}\nabla_{\mu}\nabla_{\nu}s - \frac{k}{s}g_{\mu\nu}\nabla^{\alpha}\nabla_{\alpha} s$, over the $\rho$ term.
As the universe expands, the various objects and particles move away from each other. The number of objects that can influence or interact with a given particle (star or galaxy etc),  within a time interval, $\Delta t$, decreases. 
Since $k$ depends on the number of particles or objects interacting with each other, $k$ can be expected to decrease. 
Consequently, $k$ during inflation would be higher than the $k$ during the radiation era, which in turn would be higher than the $k$ during the matter era.
As per the concordance model of cosmology, let us take the scale factor, $a_{eq}$, at the matter radiation equality to be $a_{eq} \approx 10^{-4}$.
Then, the matter density of the universe, $\rho$, would decrease as 
\begin{equation}
	\label{eq:fried_intro1}
	\frac{\rho_0}{\rho_{eq}} = \left (\frac{a_{eq}}{a_{0}} \right )^3 \approx 10^{-12},
\end{equation}
where, $\rho_0$ and $a_0$ are energy density and scale factor today, and $a_0=1$.
$\rho_{eq}$ is the energy density at equality.
Based on Eq.~\ref{eq:tfr1}, this then leads to the scaling of the degree of thermalization, $k$, as
\begin{equation}
	\label{eq:fried_intro2}
	\frac{k_0}{k_{eq}} = 
	\sqrt {\frac{\rho_0}{\rho_{eq}} } \approx \left (\frac{a_{eq}}{a_{0}} \right )^{3/2} = 10^{-6}.
\end{equation}
This implies that the degree of thermalization today, $k$, can be expected to be $10^{-6}$ times smaller than what it was during the period of matter radiation equality. At the same time, based on similar arguments,  $|k_{eq}| \ll |k_{inflation}|$. If we take $k_{inflation} \sim -1$, then $|k_{eq}|  \ll 1$.
Thus, in the matter era, the entropic term,
$\frac{k}{s}\nabla_{\mu}\nabla_{\nu}s - \frac{k}{s}g_{\mu\nu}\nabla^{\alpha}\nabla_{\alpha} s$,
may not play a significant role and the baryonic density, $\rho_b$, may dominate.
In Sec.~\ref{sec:matter_era}, we therefore reintroduce the baryonic mass density, $\rho_b$, during the matter era.

In this section, we also refer to the term, 
$\frac{k}{s}\nabla_{\mu}\nabla_{\nu}s - \frac{k}{s}g_{\mu\nu}\nabla^{\alpha}\nabla_{\alpha} s$, as entropic mass.
Sections~\ref{sec:entropymass} and~\ref{sec:newton} explore the behavior of a pure interaction model in the form of entropic mass. 
Dark matter is replaced by the entropic mass during the matter era.

\subsection{Massless scalar particle}
\label{sec:massless_scalar}

We now proceed to model a massless scalar particle in 5-D space-time-temperature.
But before we proceed, let us address an important aspect on interacting massless scalar particles being in equilibrium.
Normally, one would expect a thermal system to start un-equilibrated, and equilibrate over a period of time. 
If the universe were to start in an un-equilibrated state, then the 5-D framework is not applicable.

However, as an ansatz, we claim that the universe begins with an equilibrated state. 
Let us recognize that the time required for equilibration is proportional to the mean free path.
At $t \rightarrow 0$, the volume of the universe $\rightarrow 0$. Hence, mean free path $\rightarrow 0$, Thus, the time taken for equilibration should be instantaneous, or alternatively, the system (universe) comes up in a state of equilibrium. 
In Sec.~\ref{sec:inflation_era}, Eqs.~\ref{eq:lval} and ~\ref{eq:mval}, and the discussions related to the solution space of these two equations suggest that the universe is indeed in an equlibrated state. 
\subsubsection{Euler Lagrange equations of motion}
\label{sec:EL_equation_of_motion}
Let us consider the 5-D metric
\begin{equation}
	\label{eq:gfried_5D}
		g_{ab} = diag[-c^2, a^2(t), a^2(t), a^2(t), s^2(t)].
\end{equation}
The Lagrangian for a massless scalar is,
\begin{equation}
\label{eq:scalar_lagrangian}
	L =  \frac{1}{2}D^{\mu}\hat{\phi}^*D_{\mu} \hat{\phi}.
\end{equation}
This leads to the Klein Gordon equation of motion for a massless scalar particle:
\begin{equation}
	\label{eq:scalar_KG}
	g^{ab}D_aD_b \hat{\phi} = 0,
\end{equation}
where $\hat{\phi}$ is of the form, 
\begin{equation}
	\label{eq:scalar_phi}
	\hat{\phi} = \hat{b}^{\dagger}_{{\bf p}=0,\omega} \exp\{i(-c^2\sE(t) + s(t)^2\omega_n \beta)\},
\end{equation}
	and $\hat{b}^{\dagger}_{{\bf p},\omega}$ is the 5-D creation operator. For a homogeneous field, ${\bf p}=0$. For a gist of the 5-D annihilation and creation operator, please refer to Eqs.~\ref{eq:operator5D},~\ref{eq:commutation5D}, and~\ref{eq:pcreation}. For further details, one may refer to Ref.~\cite{gans7}. 
	The Matsubara frequency, $\omega_n$, is given by:
\begin{equation}
\label{eq:define_omega}
	\omega_n = \frac{2n\pi}{\beta_c},
\end{equation}
for integer $n$. $\beta_c$ is the inverse temperature, $\beta$, at some reference point of time, $t$, and is a constant. 
Equation~\ref{eq:scalar_KG} then gives,
\begin{equation}
	a s^2 \frac{\partial^2 \hat{\phi}}{\partial t^2} + \left \{ a(t) s(t) \frac{\partial s}{\partial t} + 3s^2\frac{\partial a}{\partial t} \right \} \frac{\partial \hat{\phi}}{\partial t} - c^2a\frac{\partial^2 \hat{\phi}}{\partial \beta^2} = 0.
\end{equation}
It is implicit that $a$ and $s$ are functions of $t$.
An external potential $V(\hat{\phi})$ is often used in literature to stimulate an expanding universe~\citep{potential1, potential2}.
MBG theory doesn't require an external potential, $V(\hat(\phi)$, since the $5^{th}$ dimension captures interactions.
Expanding $\phi$ and evaluating at $\beta = \beta_c$, we get,
\begin{multline}
\label{eq:scalar_complex}
\Big \{ \left (4 \beta_c^2 as^3 \dot{s}^2 - c^2 a s^3 \right )\omega_n^2\\
	+ \Big ( -2i\beta_c as^2 \ddot{s} - 4i\beta_c a s \dot{s}^2 + \left [-6i\beta s^2 \dot{a} - 4\dot{\sE} \beta_c c^2 a s^2 \right ] \dot{s} \Big ) \omega_n\\
	+ i\dot{\sE} c^2a \dot{s} + 3i\dot{\sE} c^2s\dot{a}  + ic^2as\ddot{\sE} + \dot{\sE}^2c^4as \Big \}\hat{\phi} = 0.
\end{multline}
The "dot" refers to derivative w.r.t. time.
The real part of Eq.~\ref{eq:scalar_complex}, when solved leads to:
\begin{equation}
\label{eq:scalar_omega_s}
	s\omega_n = \frac{c^2\dot{\sE}} {2\dot{\beta}(t) \mp c}.
\end{equation}
where, we have used Eq.~\ref{eq:define_s}, i.e., $\beta({\bf x},t) = s({\bf x},t)\beta_c$. Since we are considering a homogeneous universe, $s$ is now only a function of time.

At the beginning of time, it is reasonable to believe that either the temperature falls from an infinitely large value or increases rapidly to a large value. Then, the rate of change of inverse temperature, $|\dot{\beta}(t)|$, is very large, i.e., $|\dot{\beta}(t)| \gg \frac{c}{2}$.
We shall further discuss the temperature at the beginning of time at the end of Sec.~\ref{sec:inflation_era}.
In this limit, Eq.~\ref{eq:scalar_omega_s} becomes,
\begin{equation}
\label{eq:scalar_omega_s2}
	s\omega_n \approx \frac{c^2\dot{\sE}} {2\dot{\beta}(t) }.
\end{equation}

From Eq.~\ref{eq:scalar_phi}, and using Eq.~\ref{eq:define_s}, we get
\begin{equation}
	\partial_t \hat{\phi} = i\left \{-c^2\dot{\sE}(t) + 2s(t)\omega_n \dot{\beta}(t) \right \} \hat{\phi}.
\end{equation}
Substituting $s\omega_n$ from Eq.~\ref{eq:scalar_omega_s2}, we get
\begin{equation}
	\label{eq:scalar_slow_roll}
	\partial_t \hat{\phi} = 0.
\end{equation}
Thus, if at $t \sim 0+$, $|\dot{\beta}| \gg c$, the slow roll condition, Eq.~\ref{eq:scalar_slow_roll}, naturally follows from the Euler Lagrange equations of motion.

By substituting Eq.~\ref{eq:scalar_omega_s} into Eq.~\ref{eq:scalar_complex} and solving for the imaginary part, we come up with:
\begin{equation}
	\label{eq:scalar_imag}
	\frac{2\ddot{\beta}}{2\dot{\beta} \mp c} 
	+\left ( \frac{2\dot{\beta} \pm c}{2\dot{\beta} \mp c} \right ) \frac{\dot{\beta}}{\beta} 
	+ \left (\frac{\pm 3c}{2\dot{\beta} \mp c} \right ) \frac{\dot{a}}{a} = \frac{\ddot{\sE}}{\dot{\sE}}.
\end{equation}
Again, if at $t \rightarrow 0+$, $|\dot{\beta}| \gg \frac{c}{2}$, then Eq.~\ref{eq:scalar_imag} simplifies to:
\begin{equation}
	\label{eq:scalar_imag_2}
	\frac{\ddot{\beta}}{\dot{\beta}}
	+\frac{\dot{\beta}}{\beta} = \frac{\ddot{\sE}}{\dot{\sE}},
\end{equation}
or equivalently,
\begin{equation}
	\label{eq:scalar_imag_3}
	\frac{\ddot{s}}{\dot{s}}
	+\frac{\dot{s}}{s} = \frac{\ddot{\sE}}{\dot{\sE}},
\end{equation}
Let,
\begin{equation}
	\label{eq:scalar_sam}
	s = s_0 a^m, 
\end{equation}
where $m$ and $s_0$ are constants. 
While $s_0$ is a positive constant, we defer the discussion on the sign of $m$ to the end of Sec.~\ref{sec:inflation_era}.
Substituting $s \propto a^m$ in Eq.~\ref{eq:scalar_imag_3},
\begin{equation}
	\label{eq:scalar_imag_4}
	\frac{\ddot{a}}{\dot{a}}
	+(2m-1)\frac{\dot{a}}{a} = \frac{\ddot{\sE}}{\dot{\sE}}.
\end{equation}
Let $E = \frac{\partial \sE}{\partial t}$. This gives the traditional energy if the energy, $E$, is a constant. Then, Eq.~\ref{eq:scalar_imag_4} can be rewritten as:
\begin{equation}
	\label{eq:scalar_imag_5}
	\frac{\ddot{a}}{\dot{a}}
	+(2m-1)\frac{\dot{a}}{a} = \frac{\dot{E}}{E}.
\end{equation}
Equation~\ref{eq:scalar_imag_5} seems to suggest that the rate of increase in energy, $E$, drives the expansion of the universe. 
This inference appears to be intuitive.

\subsubsection{Scalar field stress energy tensor in 5D}
\label{sec:stress_energy_tensor}
From Eq.~\ref{eq:scalar_lagrangian} we get,
\begin{equation}
	T^1_{ab} =  -2\frac{\delta (\sqrt{g^{(5)}} L)}{\delta g^{ab}} =
	-\partial_{a}\phi^*\partial_{b}\phi + g_{ab}L. 
\end{equation}
This gives,
\begin{center}
\begin{eqnarray}
	\label{eq:fried_T1}
	T^1_{00} = -\frac{1}{2}\left (|\partial_t \phi |^2 + \frac{c^2}{s^2}|\partial_{\beta} \phi|^2 \right );\\
	T^1_{ii} = -\frac{1}{2}\left ( \frac{a^2}{c^2} |\partial_t \phi |^2 - \frac{a^2}{s^2} |\partial_{\beta} \phi|^2 \right ),~~\forall i=1,2,3;\\
	T^1_{44} = -\frac{1}{2} \frac{s^2}{c^2} \left (|\partial_t \phi |^2 + \frac{c^2}{s^2}|\partial_{\beta} \phi|^2 \right ).
\end{eqnarray}
\end{center}
The non-diagonal elements have a value of 0.
For the slow roll condition, $\partial_t \phi \sim 0$ (Eq.~\ref{eq:scalar_slow_roll} ), we get,
	\begin{center}
\begin{eqnarray}
	\label{eq:fried_T1_slowroll}
	T^1_{00} = -\frac{1}{2}\left (\frac{c^2}{s^2}|\partial_{\beta} \phi|^2 \right );\\
	T^1_{ii} = \frac{1}{2}\frac{a^2}{s^2} |\partial_{\beta} \phi|^2 ,~~\forall i=1,2,3;\\
	T^1_{44} = -\frac{1}{2} |\partial_{\beta} \phi|^2.
\end{eqnarray}
	\end{center}
In Eq.~\ref{eq:define_omega}, various $\omega_n$ were seen to be the various Matsubara eigenstates. Thus, $\partial_t \phi$ and $\partial_{\beta} \phi$, would correspond to various Matsubara eigenstates. However, while incorporating it into the MBG equations, we implicitly use the expected value.
Based on Eq.~\ref{eq:fried_T1_slowroll}, we can define the 4-D space-time  sub-tensor $T^1_{\mu\nu}$ of $T^1_{ab}$ as:
\begin{equation}
	T^1_{\mu\nu} = \frac{1}{2s^2}g_{\mu\nu} | \partial_{\beta} \phi|^2,~~\forall~\mu,\nu = 0,1,2,3.
\end{equation}

To solve the MBG equations, we need to fully determine $T^p_{ab}$ in Eq.~\ref{eq:lincomb}. We have so far determined $T^1_{\mu\nu}$, and $P_1$ in Eq.~\ref{eq:lincomb} is now replaced by $T^1_{44}$. This leaves $T^2_{\mu\nu}$ to be determined.
$T^2_{\mu\nu}$ is the non-interacting state of $\hat{\phi}$, which would be given by the $n=0$ Matsubara eigenstate in Eq.~\ref{eq:define_omega}.
For this eigenstate, $T^2_{\mu\nu}$ vanishes since $\omega_{n=0} = 0$. Then, 
\begin{equation}
	\label{eq:scalar_T2}
	T^2_{\mu\nu} = 0.
\end{equation}
All the necessary ingredients have now been obtained to solve the MBG equations during the early universe.

\subsection{The inflation era}
\label{sec:inflation_era}
In the previous section, we have shown that, in the context of a 5-D space-time-temperature, the equations of motion of a massless scalar field can lead to a slow roll conditions if the rate of change of inverse temperature, $\dot{\beta}$, is high. 
Within the framework of the MBG equations, we now explore the inflationary universe under slow roll conditions.

In Eqs.~\ref{eq:5Dmetric},~\ref{eq:ricci5D},~\ref{eq:EFE4D},~\ref{eq:EFE5D_2},
 we had expressed the 5-D Einstein field equations in terms of 4-D space-time operators (for details, one may refer Refs.~\cite{gans7, gans8}).
Due to the projection onto a 4-D space-time we need to consider only the 4-D part of the metric in Eq.~\ref{eq:gfried_5D}.
In essence, we arrive at the FLRW metric.
\begin{equation}
\label{eq:fried1}
dS^2 = -c^2dt^2 + a^2(t) \left (dx^2 + dy^2 + dz^2 \right ).
\end{equation}

We then have:
\begin{equation}
\label{eq:fried2}
\nabla^{\alpha}\nabla_{\alpha} s = - \left [ \frac{a\frac{\partial^2s}{\partial t^2} + 3\frac{\partial a}{\partial t} \frac{\partial s}{\partial t}}{c^2a} \right ],
\end{equation}
\begin{equation}
\label{eq:fried3}
\nabla_0\nabla_0 s = \frac{\partial^2 s}{\partial t^2},
\end{equation}
\begin{equation}
\label{eq:fried4}
\nabla_1 \nabla_1 s = \nabla_2\nabla_2 s = \nabla_3\nabla_3 s = -\left [\frac{a\frac{\partial a}{\partial t} \frac{\partial s}{\partial t}}{c^2} \right ],
\end{equation}
\begin{equation}
\label{eq:fried5}
\nabla_{\mu}\nabla_{\nu} s = 0 ~~~~~\forall \mu \neq \nu.
\end{equation}

The Ricci tensor would be,
\begin{equation}
\label{eq:fried_ric1}
	R_{00} = -3\frac{1}{a}\frac{\partial^2a}{\partial t^2},
\end{equation}
\begin{equation}
\label{eq:fried_ric2}
	R_{11} = R_{22} = R_{33} = \frac{1}{c^2} \left [ a\frac{\partial^2a}{\partial t^2} + 2\left( \frac{\partial a}{\partial t} \right )^2  \right ].
\end{equation}
The Ricci scalar is,
\begin{equation}
\label{eq:fried_ricci_scalar}
	R = \frac{6}{c^2 a^2}\left [ a \frac{\partial ^2 a}{\partial t^2} + \left ( \frac{\partial a}{\partial t} \right )^2 \right ].
\end{equation}

Let us frame the MBG equations from Ref.~\cite{gans8}, and also mentioned in Eq.~\ref{eq:EFE5D_p},
for the current context, using the results of Sec.\ref{sec:massless_scalar}.
\begin{eqnarray}
\label{eq:fried_mbg1a}
	-\frac{1}{2}ks^2R = \frac{8\pi G}{c^4} k s^2 T^1_{44} 
	= -\left ( \frac{1}{2} \right ) \frac{8\pi G}{c^4} k (|\partial_{\beta}\phi|^2).
\end{eqnarray}
\begin{multline}
\label{eq:fried_mbg1b}
	R_{\mu\nu} - \frac{1}{2}g_{\mu\nu} R \\
	= \frac{8\pi G}{c^4}\left \{ (1-k)T^2_{\mu\nu} + kT^1_{\mu\nu} \right \} + \frac{k}{s}\nabla_{\mu}\nabla_{\nu} s - \frac{k}{s}g_{\mu\nu}\nabla^{\alpha}\nabla_{\alpha} s.
\end{multline}

Based on Eq.~\ref{eq:scalar_T2}, $T^2_{\mu\nu}$ vanishes from Eq.~\ref{eq:fried_mbg1b}.
Substituting $\partial_{\beta}\phi$ in terms of $R$ from Eq.~\ref{eq:fried_mbg1a}, 
 and after some algebra we get,
\begin{multline}
	\label{eq:fried_mbg3}
	R_{\mu\nu} - R\frac{k+1}{2} g_{\mu\nu} 
	= \frac{k}{s}\nabla_{\mu}\nabla_{\nu}s - \frac{k}{s}g_{\mu\nu} \nabla^{\alpha}\nabla_{\alpha} s.
\end{multline}
As can be seen in Eq.~\ref{eq:fried_mbg3}, the gravitational dynamics are driven by pure interaction (or entropic) terms at the beginning of time. This confirms to the discussion in Sec.~\ref{sec:single_particle_MBG}.
Equation~\ref{eq:fried_mbg3} becomes
\begin{equation}
	\label{eq:fried_mbg4}
	R_{00} + \frac{k+1}{2}Rc^2 = \frac{k}{s}\nabla_0\nabla_0 s + \frac{kc^2}{s}\nabla^{\alpha}\nabla_{\alpha} s,
\end{equation}
and,
\begin{equation}
	\label{eq:fried_mbg5}
	R_{11} - \frac{k+1}{2}Ra^2 = \frac{k}{s}\nabla_1\nabla_1 s - \frac{ka^2}{s}\nabla^{\alpha}\nabla_{\alpha} s.
\end{equation}
Substituting various quantities from Eq.~\ref{eq:fried2} to Eq.~\ref{eq:fried_ricci_scalar} in Eqs.~\ref{eq:fried_mbg4} and~\ref{eq:fried_mbg5} we get,
\begin{equation}
	\label{eq:fried_mbg6}
	 \left ( \frac{\dot{a}}{a} \right )^2(1+k) = -k\left [ \frac{\ddot{a}}{a} + \frac{\dot{a}}{a}\frac{\dot{s}}{s} \right ],
\end{equation}
\begin{equation}
	\label{eq:fried_mbg7}
	(-2-3k)\frac{\ddot{a}}{a} + (-1 -3k)\left ( \frac{\dot{a}}{a} \right )^2 = 2k\frac{\dot{a}}{a}\frac{\dot{s}}{s} + k\frac{\ddot{s}}{s}.
\end{equation}
Substituting Eq.~\ref{eq:define_s} in Eq.~\ref{eq:fried_mbg6} and~\ref{eq:fried_mbg7}, and simplifying we get respectively,
\begin{equation}
	\label{eq:fried_mbg9}
	(1 + k + km) \left (\frac{\dot{a}}{a} \right )^2 = -k \frac{\ddot{a}}{a};
\end{equation}
\begin{equation}
	\label{eq:fried_mbg10}
	(-1 - 3k - km^2 - km) \left (\frac{\dot{a}}{a} \right )^2
	= (km + 2 + 3k) \frac{\ddot{a}}{a}.
\end{equation}
For an exponential inflation of the kind $a \propto \exp(-Ht)$,
	$\frac{\ddot{a}}{a} =  \left (\frac{\dot{a}}{a} \right )^2$. Then, solving Eqs.~\ref{eq:fried_mbg9} and \ref{eq:fried_mbg10}, we obtain two possible solutions: 
\begin{itemize}
\item $k=-\frac{1}{3}, m = 1$. 
\item $k=-\frac{1}{2}, m = 0$. 
\end{itemize}

\begin{figure}
\includegraphics[width = 80mm,height = 80mm]{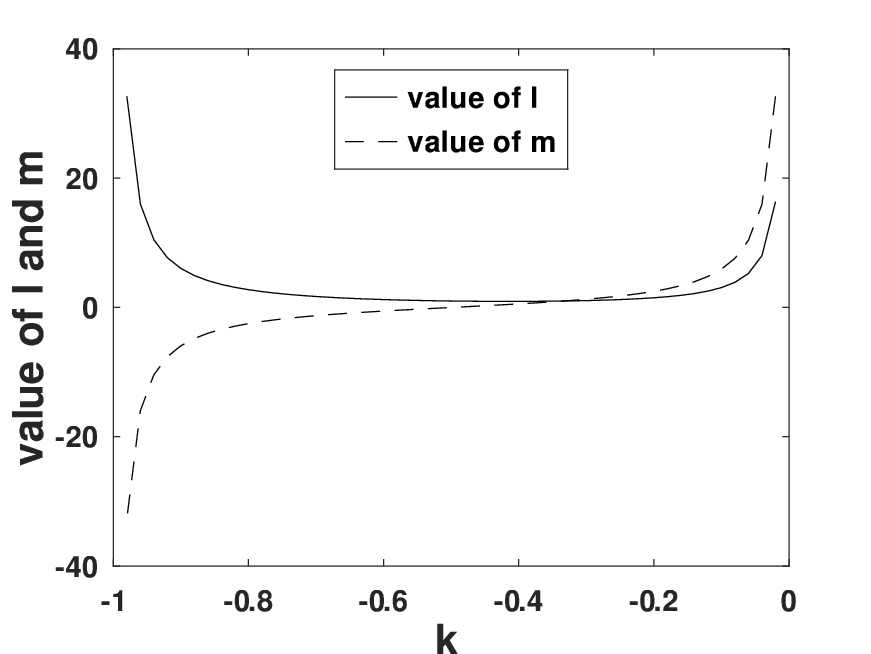}
	\caption{Relation between $l$ and $k$.}
\label{fig:kl}
\end{figure}

However, there is no particular reason why inflation should be restricted to  exponential acceleration.
For an inflation that is more accelerated than exponential, we take
\begin{equation}
	\frac{\ddot{a}}{a} =  l\left (\frac{\dot{a}}{a} \right )^2,~~~ l> 1.
\end{equation}
Solving Eqs.~\ref{eq:fried_mbg9} and \ref{eq:fried_mbg10}, we obtain.
\begin{equation}
	\label{eq:lval}
	l = -\frac{1}{3}\frac{\left ( 3k^2 + 2k + 1 \right )} {\left (k+1 \right )k},
\end{equation}
and,
\begin{equation}
	\label{eq:mval}
	m = -\frac{1}{k} \left ( 1 + k + kl \right ).
\end{equation}
From, Eq.~\ref{eq:lval}, it is obvious that any positive value of $k$, would lead to a negative $l$. A negative $l$ means $\ddot{a} < 0$ and will not lead to an accelerated universe. Thus, a negative $k$ is required for an accelerated expansion of the universe and this aligns with the analysis in Ref.~\cite{gans8E}.
Fig.~\ref{fig:kl} shows the relationship between $l$ v/s $k$ and $m$ v/s $k$.
For a significant range of $k$, $l$ is close to 1. This indicates that the inflation would be close to exponential for quite a significant range of $k$. However, if $k ~ -1$, $l$ explodes. 
It's conceivable that the system was fully thermalized with $k=-1$ at the beginning of the Big Bang singularity.
The universe expanded with an acceleration that was much higher than an exponential.
It is probable that this infinitely huge accelerated expansion led to the creation of the universe from nothingness.
Subsequently, as the value of $|k|$ becomes smaller, the exponential inflation period began, until a phase change led to the end of inflation.
However, when $k < -0.5$, $m$ is negative. This means that the inverse temperature decreases, or equivalently, temperature increases. 
Thus, Fig.~\ref{fig:kl} might have the following interpretations:
\begin{enumerate}
\item The domain in Fig.~\ref{fig:kl}, where $m>0$ is the only valid region, i.e. the domain, $k> -0.5$. 
This means that the singularity initiates with an infinitely high temperature and then cools down.
\item The entire domain in Fig.~\ref{fig:kl}, i.e., $0>k>-1$, is valid. This opens up the possibility that the Big Bang singularity started from nothingness, with the temperature, $\Theta \rightarrow 0$. 
As energy, $E$, accumulates, the temperature rises rapidly.
The rate at which the universe expands is beyond exponential.
The temperature can increase in an expanding universe if energy or scalar field creation occurs at a sufficiently rapid pace, as governed by Eq.~\ref{eq:scalar_imag_2}.
At this moment, $k \sim -1$. Subsequently, there is a reversal and the temperature begins to cool, while the universe continues to expand at a roughly exponential rate.
\end{enumerate}
It appears that the second scenario is more likely than the first because it is improbable for nothingness to transform into an infinite temperature instantly.
But then again, the $5^{th}$ temperature dimension is a topological cylinder~\citep{gans6,gans5}, with  $\beta = 0$,  and $\beta= \beta_c$ being the same point. Hence, the former case cannot be ruled out either.
More investigation would be required into these aspects.

Referring to Eq.~\ref{eq:scalar_imag_5}, a negative $m$ implies that the ratio $\frac{\ddot{a}}{\dot{a}}$ has to be significantly be higher than $\frac{\dot{a}}{a}$ to maintain positive increase in energy, $E$. This again leads to an explosively accelerated expanding universe. 
It might be worth noting that the theory of MBG opens up new possibilities for inflation beyond what exists with Einstein's gravity.

\subsection{Entropic mass hierarchy}
\label{sec:entropymass}

In Eq.~\ref{eq:scalar_sam},
we have taken $s(t)\propto a(t)^m$, which implies that $s$ is solely dependent on time. However, in Refs.~\cite{gans8,gans9}, it was seen that, for a galaxy or a galaxy cluster, $s$ was purely a function of radial distance, $r$.
Similarly, the local metric at a localized region of space-time (for example a galaxy) need not be the same as the metric describing the Friedmann universe. The same argument would apply to the degree of thermalization, $k$. 
This is so because the interaction between the stars within a galactic system can be different from say the interaction between galaxy clusters in the universe.
Thus, if we wish to model a local object like a galaxy in the cosmic framework, we would need a mechanism to model two different $s$, $g_{\mu\nu}$ and $k$. 
One, for the local object, such as a galaxy, and the second for the cosmos modeled by the FLRW metric. This gives rise to an entropy mass hierarchy, with the entropic mass at each level determined by the variation of $s$, the metric, $g_{\mu\nu}$ and the degree of thermalization, $k$, at that level.
Let us label the variation in inverse temperature, $s$, $g_{\mu\nu}$ and $k$ for  the local body as $s_l$, $g^l_{\mu\nu}$, $k_l$, and retain $s$, $g_{\mu\nu}$ and $k$ for the Friedmann cosmos. The index $l$ is a superscript for $g^l_{\mu\nu}$ for ease of representation only, and has no other significance.
We would need to model the thermal or entropic part of the local body (galaxy) as a particle, with a stress energy tensor having trace $\rho^E_0$ and a hypothetical four velocity $v^h_{\mu}$. We shall soon explain why $v^h_{\mu}$ is a hypothetical 4-velocity. 
To this effect, let us start with the MBG equations:
\begin{equation}
	\label{eq:ten1}
	R^l_{\mu\nu} - \frac{k_l}{s_l}\nabla_{\mu}\nabla_{\nu}s_l - \frac{1}{2}g^l_{\mu\nu}R^{l\,(4)} + \frac{k_lg^l_{\mu\nu}}{s_l}\nabla^{\alpha}\nabla_{\alpha}s_l 
	= \frac{8\pi G}{c^4} T_{\mu\nu},
\end{equation}
where, 
\begin{multline}
	\label{eq:ten2}
	T_{\mu\nu} = \left (\rho + \frac{1}{c^2} (1-k_l)P_2 + k_lP_1 ) u_{\mu}u_{\nu} \right )\\
	+ g_{\mu\nu}\left ( (1-k_l)P_2 + k_lP_1 \right ),
\end{multline}
and
\begin{equation}
	\label{eq:ten3}
	-\frac{1}{2} k_ls_l^2 R^l = \frac{8\pi G}{c^4} k_lP_1 s_l^2.
\end{equation}
Contracting Eq.~\ref{eq:ten1}, we get
\begin{equation}
	\label{eq:ten4}
	R^l(1 - \frac{D}{2}) + \frac{k_l}{s_l}(D-1)\nabla^{\alpha}\nabla_{\alpha} s_l = \frac{8\pi G}{c^4}T,
\end{equation}
where, the number of dimensions, $= D =4$
The entropic or thermal component comes from, 
$\frac{k_l}{s_l}(D-1)\nabla^{\alpha}\nabla_{\alpha} s_l = -\frac{8\pi G}{c^4}T^E$ (say). We wish to find a 4-vector $w_{\mu}$ such that,  
\begin{equation}
	\label{eq:ten5}
	-\frac{8\pi G}{c^4}T^E w_{\mu}w_{\nu} = -\frac{k_l}{s_l}\nabla_{\mu}\nabla_{\nu} s_l + \frac{k_l}{s_l} g^l_{\mu\nu} \nabla^{\alpha}\nabla_{\alpha} s_l,
\end{equation}
We note that $w_{\mu}$ is only a mathematical object, and is not a physical 4-velocity.
Substituting $T^E$ in Eq.~\ref{eq:ten5},
\begin{equation}
	\label{eq:ten6}
	\frac{k_l}{s_l} (D-1) \nabla^{\alpha}\nabla_{\alpha}s_l w_{\mu}w_{\nu} = -\frac{k_l}{s_l}\nabla_{\mu}\nabla_{\nu} s_l + \frac{k_l}{s_l} g^l_{\mu\nu} \nabla^{\alpha}\nabla_{\alpha} s_l.
\end{equation}
We can proceed a little bit further, if for the local body we use a spherically symmetric metric, namely,
\begin{equation}
	\label{eq:ten_metric}
	g^l_{\mu\nu} = 
                \left[ \begin{array}{c c c c }
			-c^2 & 0 & 0 & 0\\
			0 & 1 & 0 & 0\\
			0 & 0 & r^2 & 0\\
			0 & 0 & 0 & r^2 sin^2(\theta)\\
                \end{array} \right ].
\end{equation}
The local space-time metric, $g^l_{\mu\nu}$, is only applicable to local bodies such as galaxies. The global metric describing the cosmos is the FLRW metric mentioned in Eq.~\ref{eq:fried1}.
In this case, Eq.~\ref{eq:ten6} can be written as the set of equations:
\begin{eqnarray}
	\label{eq:ten_8}
	(D-1)\nabla^{\alpha}\nabla_{\alpha}s_l\, w_0 w_0 = -c^2\nabla^{\alpha}\nabla_{\alpha} s_l,\\
	(D-1)\nabla^{\alpha}\nabla_{\alpha}s_l\, w_1 w_1 = -\frac{\partial^2s_l}{\partial r^2} + \nabla^{\alpha}\nabla_{\alpha} s_l,\\
	(D-1)\nabla^{\alpha}\nabla_{\alpha}s_l\, w_2 w_2 = -r\frac{\partial s_l}{\partial r} + r^2\nabla^{\alpha}\nabla_{\alpha} s_l,\\
\nonumber	(D-1)\nabla^{\alpha}\nabla_{\alpha}s_l\, w_3 w_3 =\\
	-r \sin^2(\theta) \frac{\partial s_l}{\partial r} + r^2 \sin^2(\theta)\nabla^{\alpha}\nabla_{\alpha} s_l.
\end{eqnarray}
Solving, we get,
\begin{equation}
	\label{eq:ten_13}
	w_0 = \frac{ic}{\sqrt{D-1}},
\end{equation}
\begin{equation}
	\label{eq:ten_14}
	w_1 = \sqrt{ \frac{ \frac{ 2}{r} \frac{\partial s_l}{\partial r} } {(D-1)\nabla^{\alpha}\nabla_{\alpha} s_l} },
\end{equation}
\begin{equation}
	\label{eq:ten_15}
	w_2 = \sqrt{ \frac{ r \frac{\partial s_l}{\partial r} + r^2\frac{\partial^2 s_l}{\partial r^2} } {(D-1)\nabla^{\alpha}\nabla_{\alpha} s_l} }.
\end{equation}
and
\begin{equation}
	\label{eq:ten_16}
	w_3 = \sin(\theta) \sqrt{ \frac{ r \frac{\partial s_l}{\partial r} + r^2\frac{\partial^2 s_l}{\partial r^2} } {(D-1)\nabla^{\alpha}\nabla_{\alpha} s_l} }.
\end{equation}
It can be seen that,
\begin{equation}
	\label{eq:ten_17}
	w^{\mu}w_{\mu} = \frac{1}{D-1}.
\end{equation}
The norm of the $w_{\mu}$ vector in Eq.~\ref{eq:ten_17}, makes it clear why $w_{\mu}$ is not a 4-velocity. 
To maintain compatibility with the usual definition of a four velocity, let us define 
\begin{equation}
	\label{eq:ten_18}
	v^h_{\mu} = ic\sqrt{D-1} w_{\mu}.
\end{equation}
The superscript $h$ stands for "hypothetical". There is no physical velocity associated with the $\frac{k_l}{s_l}(D-1)\nabla^{\alpha}\nabla_{\alpha} s_l$  term. As a result, $v_{\mu}^h$ is being referred to as a hypothetical 4-velocity.
Then, we get,
\begin{equation} 
	\label{eq:ten_19}
	v^{h\,\mu}v^h_{\mu} = -c^2,
\end{equation} 
which is the expected inner product for the metric in Eq.~\ref{eq:ten_metric}.
Rewrite Eq.~\ref{eq:ten6} as:
\begin{equation}
	\label{eq:ten20}
	-\frac{k_l}{s_lc^2} \nabla^{\alpha}\nabla_{\alpha} s_l v^h_{\mu}v^h_{\nu} = -\frac{k_l}{s_l}\nabla_{\mu}\nabla_{\nu} s_l + \frac{k_l}{s_l} g^l_{\mu\nu} \nabla^{\alpha}\nabla_{\alpha} s_l,
\end{equation}
and define $\rho_0^E$ as:
\begin{equation}
	\label{eq:ten21}
	\frac{8\pi G}{c^4}\rho_0^E = -\frac{k_l}{s_lc^2} \nabla^{\alpha}\nabla_{\alpha} s_l.
\end{equation}
This finally yields,
\begin{equation}
	\label{eq:ten22}
	\rho_0^E = -\frac{k_lc^2}{8\pi G }\frac{1}{s_l} \nabla^{\alpha}\nabla_{\alpha} s_l.
\end{equation}
In Ref.~\cite{gans8E}, it was shown that $k_l <0$, and also the subsequent interpretation of a negative $k_l$ was discussed. With $k_l<0$, $\rho_0^E$ is positive for positive $\nabla^{\alpha}\nabla_{\alpha}s_l$.
We can now call $\rho^E_0$, the entropy mass density. After substituting $s_l \rightarrow \frac{1}{\phi}$, this can be seen identical to the pseudo mass expression in~\citep{gans8}, which impacted the galaxy rotation curves.

Against the backdrop of the formulation in Ref.~\cite{gans8,gans9},
the above discussion also highlights a crucial fact. 
Galaxy rotation curves primarily rely on $\rho^E_0 v^h_0 v^h_0$, while gravitational lensing relies on both $\rho^E_0 v^h_0 v^h_0$ and $\rho^E_0 v^h_1v^h_1$.
If the entropic term in MBG is equivalent to dark matter in $\Lambda CDM$ cosmology, then
the amount of dark matter inferred from the rotation curves and the gravitational lensing should be different.
In fact, the literature does contain references to such scenarios~\citep{rot_lens}.
Another example is that the dark matter inferred via gravitational lensing for NGC6505 is about 11.1\%~\citep{ngc6505}, while the dark matter inferred via rotation curves for the Milkyway is more than 90\%~\citep{sofue_mw}.
In the MBG theory, the galaxy rotation curve and the gravitational lensing are seen as arising from different components of a tensor.

\subsection{Newtonian behavior of $\rho^E_0$ for weak fields at far distances}
\label{sec:newton}
In Ref.~\cite{gans8,gans9}, it was shown that $s\propto \frac{1}{\phi}$ for systems such as a galaxy. 
This led to $\frac{kc^2}{2s}\nabla^{\alpha}\nabla_{\alpha}s \rightarrow \frac{kc^2}{2}\phi\nabla^{\alpha}\nabla_{\alpha} \frac{1}{\phi}$ for such systems.
With this representation, we can examine the behavior of $\rho^E_0$.
For a purely interacting system ($\rho \rightarrow 0$), after replacing $k$ by $k_l$, the MBG equation, Eq.~\ref{eq:gal10}, becomes:
\begin{equation}
	\label{eq:newton1}
	\nabla^2 \phi \approx  -k_lc^2\frac{\phi}{2}\nabla^2\frac{1}{\phi}.
\end{equation}
We note that Eq.~\ref{eq:newton1} is of the form:
\begin{equation}
	\label{eq:newton1b}
   	\nabla^2 \phi \approx 4\pi G\rho^E_0,
\end{equation}
which can be seen as Newton's equation of gravity with entropic mass density, $\rho^E_0$.
At far distances, a system such as a galaxy can be treated as a point mass. 
This results in a spherical symmetry.
Under spherical symmetry, the Eq.~\ref{eq:newton1} can be written as:
\begin{equation}
	\label{eq:newton2}
	        \frac{1}{r^2}\frac{\partial}{\partial r} r^2 \frac{\partial \phi}{\partial r} \approx K\frac{\phi}{r^2}\frac{\partial}{\partial r} r^2 \frac{\partial}{\partial r}\frac{1}{\phi},
\end{equation}
where,
\begin{equation}
	\label{eq:newton2b}
	K = -\frac{k_lc^2}{2}.
\end{equation}

The solution to the equation mentioned above was derived in Ref.~\cite{gans8}. For ease of reference, the solution is reproduced in the Appendix$^1$.\footnotetext[1]{Typographical errors in the solution are also corrected.} From Eq.~\ref{eq:asym8}, the solution to the above equation is:
\begin{equation}
\label{eq:newton3}
	\frac{K^2}{\phi} - 2K\phi\ln(|\phi|) - \phi  =  \frac{c_1}{r} + c_2,
\end{equation}
where, $c_1$ and $c_2$ are constants of integration. 

It's reasonable to assume that $\phi$ will become very small at a large $r$.
This means, $c_2 \ne 0$, due to the $\frac{K^2}{\phi}$ term. In the limit of small $\phi$ we have,
\begin{equation}
	\label{eq:newton4}
	\lim_{\phi \rightarrow 0} \phi \ln(|\phi|) = 0.
\end{equation}
Then, $\frac{K^2}{\phi} - 2K\phi\ln(|\phi|) - \phi \approx \frac{K^2}{\phi}$. 
Subsequently, Eq.~\ref{eq:newton3} becomes:
\begin{equation}
\label{eq:newton5}
	\frac{K^2}{\phi}  \approx  \frac{c_1}{r} + c_2,
\end{equation}
Differentiating and rearranging,
\begin{equation}
\label{eq:newton6}
	\frac{\partial\phi}{\partial r}   =  \frac{c_1\phi^2}{K^2r^2},
\end{equation}
Substituting $\phi$ from Eq.~\ref{eq:newton5}, 
\begin{equation}
\label{eq:newton7}
	\frac{\partial\phi}{\partial r}   =  \frac{c_1K^2}{(c_1 + c_2r)^2},
\end{equation}
For large $r$, Eq.~\ref{eq:newton7} simplifies to
\begin{equation}
\label{eq:newton8}
	\frac{\partial\phi}{\partial r}   =  \frac{c_1K^2}{c_2^2r^2}.
\end{equation}
This result indicates that at large $r$, the gravitational potential behaves like a Newtonian field with an equivalent point mass, $M_N^{eq}$, where
\begin{equation}
\label{eq:newton9}
	M_N^{eq} =  \frac{c_1K^2}{Gc_2^2},
\end{equation}
and $G$ is Newton's gravitational constant.
As a case in point, a galaxy, along with both baryonic mass, $\rho$, and  entropic mass or pseudo mass, $\rho^E_0$ (the equivalent of dark matter in the MBG theory), should behave like a particle with Newtonian gravity at distances far away from the galaxy.
Interestingly, a similar behavior is observed at the outermost edge of the Milkyway. While a flat rotation curve is seen for most of the galaxy, the velocity of stars at the very edge of the Milkyway galaxy is seen to decrease~\citep{hammer}. As a caution, the results may also depend on the methodology used to obtain the rotation curve. 
Secondly, it remains to be seen whether Milky Way is a one-off phenomenon.

Let us now investigate the relationship between $M_N$ and $k$. 
From Eq.~\ref{eq:newton1}, since the factor $\phi\nabla^2\frac{1}{\phi}$ is invariant w.r.t. scaling of $\phi$ by a constant, it can easily be seen that,
\begin{eqnarray}
\label{eq:newton10}
	\nabla^2 \phi \propto k_l,~~~~ 
\phi \propto k_l.
\end{eqnarray}
Since, $K \propto k_l$, Eqs.~\ref{eq:newton10} leads to,
\begin{equation}
\label{eq:newton12}
	\frac{K^2}{\phi} \propto k_l.
\end{equation}
Finally, from Eq.~\ref{eq:newton5}, we have
\begin{equation}
\label{eq:newton13}
	c_1, c_2 \propto \frac{K^2}{\phi} \propto k_l.
\end{equation}
Using the relations in Eqs.~\ref{eq:newton2b} and~\ref{eq:newton13} in Eq.~\ref{eq:newton9}, we obtain,
\begin{equation}
	M_N^{eq} \propto k_l,
\end{equation}
where, $M_N^{eq}$ is the equivalent Newton mass.
Since $M_N^{eq}$ is an outcome of the term $\phi\nabla^2 \frac{1}{\phi}$, which in turn is derived from $\frac{1}{s}\nabla^2 s$ (refer Eqs.~\ref{eq:gal8} and~\ref{eq:gal10}), one may infer that $\frac{1}{s}\nabla^2 s$ acts as a Newtonian particle for a time invariant weak field and at far away distances.

As $k_l\rightarrow 0$, it is seen that $M_N^{eq} \rightarrow 0$. Thus, when there are no interactions, like in the case of a single particle, there is no equivalent mass, $M_N^{eq}$. Consequently, there is no impact to time.

\subsection{Matter era }
\label{sec:matter_era}
The result of Sec.~\ref{sec:entropymass} can be utilized to model the Friedman universe deep in the matter era.
Let us start from the MBG equations, i.e. Eq~\ref{eq:fried_mbg1a}
\begin{eqnarray}
\label{eq:fried_mat1}
	-\frac{1}{2}ks^2R = \frac{8\pi G}{c^4} kP_1 s^2, 
\end{eqnarray}
\begin{eqnarray}
\label{eq:fried_mat2}
\nonumber	R_{\mu\nu} - \frac{k}{s}\nabla_{\mu}\nabla_{\nu} s - \frac{1}{2}g_{\mu\nu} R + \frac{k}{2}g_{\mu\nu}\nabla^{\alpha}\nabla_{\alpha} s = \frac{8\pi G}{c^4} T_{\mu\nu},\\
\end{eqnarray}
where,
\begin{equation}
\label{eq:fried_mat3}
	T_{\mu\nu} = \left ( \rho + \frac{kP_1}{c^2} \right ) u_{\mu}u_{\nu} + kg_{\mu\nu} P_1.
\end{equation}
Here, $k$, $s$ and $g_{\mu\nu}$ refer to the values in the Friedman universe.
But we now redefine $\rho$ as,
\begin{equation}
	\label{eq:cmb4}
	\rho = \rho_b + \rho^E_0.
\end{equation}
where, $\rho_b$ is the baryonic mass density, and $\rho_0^E$ is based on Eq.~\ref{eq:ten22}. $\rho_0^E$ refers to the entropic mass found in any local body like a galaxy, galaxy cluster etc.
As discussed in Sec.~\ref{sec:kvalues}, if much after inflation during the matter era, we were to take $k \sim 0$, then Eq.~\ref{eq:fried_mat2} would become,
\begin{eqnarray}
\label{eq:fried_mat5}
	R_{\mu\nu} - \frac{1}{2}g_{\mu\nu} R  = \frac{8\pi G}{c^4} T_{\mu\nu},
\end{eqnarray}
where,
\begin{equation}
\label{eq:fried_mat6}
	T_{\mu\nu} =  \rho u_{\mu}u_{\nu}.
\end{equation}
It is important to note that $\rho_0^E$ continues to be non-zero as $k_l \neq k$ and $k_l \neq 0$.
It is also seen that in Eq.~\ref{eq:fried_mat6}, there is no pressure.
Deep in the matter era, the pressure, $P_1 \sim 0$, as the baryonic matter is non-relativistic.
For a non-relativistic system, pressure can be neglected.
The inclusion of the cosmological constant in Eq.~\ref{eq:fried_mat5}, results in: 
\begin{eqnarray}
\label{eq:fried_mat7}
	R_{\mu\nu} - \frac{1}{2}g_{\mu\nu} R  = \frac{8\pi G}{c^4} T_{\mu\nu} - \Lambda g_{\mu\nu}.
\end{eqnarray}
This is exactly the same equation as the $\Lambda CDM$ model, with dark matter replaced by $\rho^E_0$.
The cosmological constant has been used as a placeholder because dark energy has not yet been analyzed within the MBG framework.
In Ref.~\cite{gans8}, it was shown that the pseudo mass ($\rho^E_0$ in this paper) is capable of reproducing the Navarro-Frenk-White (NFW) curve for dark matter. 
This leads credence to the claim that we can replace $\rho_d$ with $\rho^E_0$, where $\rho_d$ is the dark matter density.
Thus, in the deep matter era, if $\rho^E_0$ is taken to be of the same amount as $\rho_d$, then, MBG should reproduce to a fair extent the results of $\Lambda CDM$, for the phenomena that are confined to the matter era.

As a passing note, the modeling of the radiation era and the era close to matter radiation equality may be more intricate, as $k$ may not be taken as 0. Also, the pressure due to radiation is not zero or negligible. 
Another aspect that has not been covered is the transition from inflation to the radiation era, via reheating etc.
These require further analysis and are planned for future work.

	\section{Conclusion}
	\label{sec:conclusion}

In this work, we have shown conceptually and analytically that interaction or entropy seems to be an ingredient in the laws of gravity.
In General Relativity, for the case of cosmic inflation, interaction is captured as an external potential, $V(\hat{\phi})$. 
In the case of MBG, interaction is captured by the $5^{th}$ dimension of the massless scalar field and the entropic terms of MBG.
In previous work~\cite{gans8, gans9}, it was shown that for galaxy rotation curves and the gravitational lensing of the bullet cluster, the entropic terms of MBG are able to replace dark matter. 
This raises the question as to whether dark matter is actually the contribution of entropy to gravity.
We have shown that a massless scalar field is required for consistency with the QFT results, and fuels inflation in the MBG framework also. 
The entropic terms of MBG seem to accelerate inflation above and beyond what is scoped by General Relativity.

The theory of MBG not only reproduces cosmic inflation, but it also hints at the possibility of accelerated inflation at the beginning of inflation.
However, there are certain aspects that yet remain to be covered.
A rigorous criterion for ending inflation and the power spectrum of the curvature perturbation needs to be established. 
The matter-radiation equality and the radiation era need to be modeled, which would enable cosmic microwave background (CMB) radiation analysis.
These are planned for future work.


Data availability statement\\ 
No new data were created or analyzed in this study.

\appendix
\section{The solution of MBG equation with no mass density term}
\label{sec:tully_appendix}
For an observer far away from the system we are trying to model (say a galaxy), , one may approximate the mass to be centered at the origin. As such, 
one may assume a spherical symmetry as an approximation.
We now determine the solution to the asymptotic equation, Eq.~\ref{eq:newton1} in spherical co-ordinates. In spherical co-ordinates,  Eq.~\ref{eq:newton1} becomes
\begin{equation}
	\label{eq:gal13}
	        \frac{1}{r^2}\frac{\partial}{\partial r} r^2 \frac{\partial \phi}{\partial r} \approx -\frac{kc^2}{2}\frac{\phi}{r^2}\frac{\partial}{\partial r} r^2 \frac{\partial}{\partial r}\frac{1}{\phi}.
\end{equation}
Substitute $r = e^t$, and subsequently, assign $\zeta(\phi) = \frac{\partial \phi}{\partial t}$. This gives
\begin{equation}
\label{eq:asym2}
\zeta' - \frac{2K}{\phi^2 + K\phi}\zeta = -1,
\end{equation}
where $K = \frac{kc^2}{2}$. The integration factor for this is
\begin{equation}
\label{eq:asym3}
	f(\phi) = \left ( \frac{K + \phi}{\phi} \right )^2.
\end{equation}
Multiplying the integration factor on both sides of Eq.~\ref{eq:asym2}, and solving, we get
\begin{equation}
\label{eq:asym4}
	f(\phi) \frac{\partial \phi}{\partial t} = -\int f(\phi) d\phi.
\end{equation}
After solving, we get:
\begin{equation}
\label{eq:asym5}
	-\int f(\phi) d\phi = \frac{K^2}{\phi} - 2K\phi\ln(|\phi|) - \phi - c_2.
\end{equation}
One may again rewrite Eq.~\ref{eq:asym4} as:
\begin{equation}
\label{eq:asym6}
	\frac{f(\phi) d\phi}{\int f(\phi) d\phi} = -dt.
\end{equation}
Solving the integration and reverse substituting $t = \ln(r)$,
\begin{equation}
\label{eq:asym7}
	\int f(\phi) d\phi = -\frac{c_1}{r}.
\end{equation}
Comparing Eqs.~\ref{eq:asym5} and~\ref{eq:asym7}, we finally obtain:
\begin{equation}
\label{eq:asym8}
	\frac{K^2}{\phi} - 2K\phi\ln(|\phi|) - \phi  =  \frac{c_1}{r} + c_2,
\end{equation}
where, $c_1$ and $c_2$ are constants of integration. 


\begin{thebibliography}{99}

	\bibitem{Ein1}	Einstein, Albert 
		"The Foundation of the General Theory of Relativity" (PDF). 
		Annalen der Physik. {\bf 354} (7): 769–822. (1916). 
	\bibitem{shap1}		Shapiro, Irwin I. 
		"New Method for the Detection of Light Deflection by Solar Gravity". 
		Science, {\bf 157}, 3790, 806–808 (1967). 
	\bibitem{hol}		Holberg, J. B., 
		"Sirius B and the Measurement of the Gravitational Redshift", 
		Journal for the History of Astronomy, Vol. 41, 1, 41-64. (2010)
	\bibitem{sch}	Schiff, L. I. 
		"On Experimental Tests of the General Theory of Relativity". 
		American Journal of Physics. 28 (4): 340–343. (1960). 

	\bibitem{foma} Fomalont, E.B.; Kopeikin S.M.; Lanyi, G.; Benson, J. 
		"Progress in Measurements of the Gravitational Bending of Radio Waves Using the VLBA". 
		The Astrophysical Journal, 699 (2), 1395–1402, (2009). 
		arXiv:0904.3992. 
	\bibitem{shap2}	Shapiro, I. I. 
		"Fourth test of general relativity". 
		Physical Review Letters. {\bf 13} 26, 789–791. (1964). 
	\bibitem{ber} Bertotti B.; Iess L.; Tortora P. . 
		"A test of general relativity using radio links with the Cassini spacecraft". 
		Nature. {\bf 425}, 6956, 374–376. (2003).

	\bibitem{ligo1}		Abbott, B. P. et al.  (LIGO Scientific Collaboration and Virgo Collaboration), 
 		"ASTROPHYSICAL IMPLICATIONS OF THE BINARY BLACK HOLE MERGER GW150914",
		Phys. Rev. Lett. {\bf 116}, 061102 (2016).
	\bibitem{ligo2}		Abbott, B. P. et al., 
 		"ASTROPHYSICAL IMPLICATIONS OF THE BINARY BLACK HOLE MERGER GW150914",
The Astrophys. J. Lett. {\bf 818}, L22 (2016).
\bibitem{lens1}	A. van der Wel,  et al.,  
	"Discovery of a Quadruple Lens in CANDELS with a Record Lens Redshift z=1.53".
	Astrophysical Journal Letters. {\bf 777}, 1 L17 (2013); 
	arXiv:astro-ph.CO/1309.2826.
\bibitem{lens2} Kenneth C. Wong, et al., 
	"Discovery of a Strong Lensing Galaxy Embedded in a Cluster at z = 1.62". 
	Astrophysical Journal Letters. {\bf 789}, L31 (2014). 
	arXiv:astro-ph.GA/1405.3661.

\bibitem{rub1}	V. Rubin, Ford Jr., W. K. 
	"Rotation of the Andromeda Nebula from a Spectroscopic Survey of Emission Regions". 
	Astrophysical Journal. {\bf 159}, 379.  (1970). 
\bibitem{rub2}V. Rubin, W. K. Ford Jr, N. Thonnard, 
	"Rotational Properties of 21 Sc Galaxies with a Large Range of Luminosities and Radii from NGC 4605 (R=4kpc) to UGC 2885 (R=122kpc)". 
	Astrophysical Journal. {\bf 238}, 471, (1980). 
\bibitem{galaxy1} Y. Sofue et. al., 
	"Central Rotation Curves of Spiral Galaxies",
	The Astrophysical Journal, {\bf 523}, 136-146 (1999). 
\bibitem{galaxy2} Kyu-Hyun Chae et. al., 
	"Testing the Strong Equivalence Principle: Detection of the External Field Effect in Rotationally Supported Galaxies"
	The Astrophysical Journal, {\bf 904}, 51 (2020).
	\bibitem{dark1} Copi, C.J.; Schramm, D.N.; Turner, M.S. 
		"Big-Bang Nucleosynthesis and the Baryon Density of the Universe". 
		Science. {\bf 267} 5195, 192–199, (1995). 
		arXiv:astro-ph/9407006. 
	\bibitem{dark2} Buckley, Matthew R.; Difranzo, Anthony 
		"Collapsed Dark Matter Structures". 
		Physical Review Letters. {\bf 120}, 5, 051102, (2018). 
		arXiv:1707.03829. 
	\bibitem{dark3} Bertone, G.; Merritt, D. 
		"Dark Matter Dynamics and Indirect Detection". 
		Modern Physics Letters A. {\bf 20}, 14, 1021–1036 (2005). 
		arXiv:astro-ph/0504422. 
	\bibitem{dark4}	Overduin, J. M.; Wesson, P. S. 
		"Dark Matter and Background Light". 
		Physics Reports. {\bf 402}, 5–6, 267–406. (2004).
		arXiv:astro-ph/0407207. 
	\bibitem{dark5}	Guiot, B; Borquez, A.; Deur, A.; Werner, K. 
		"Graviballs and Dark Matter". 
		JHEP. {\bf 2020}, 11, 159, (2020). 
		arXiv:2006.02534. 
	\bibitem{mond1}	M. Milgrom, 
	"MOND theory",
Canadian Journal of Physics, {\bf 93}, 2 (2015). 
	\bibitem{mond2}. M. Milgrom. 
		"A modification of the Newtonian dynamics as a possible alternative to the hidden mass hypothesis"
		Astrophys. J. {\bf 270}, 365 (1983). 
	\bibitem{mond3} M. Milgrom. 
		"The mond limit from spacetime scale invariance",
		Astrophys. J. {\bf 698}, 1630 (2009). 
	\bibitem{mond4} M. Milgrom. 
	"Gravitational waves in bimetric MOND",
		Phys. Rev. D, {\bf 89}, 024027 (2014). 
\bibitem{mond5} 2. B. Famaey and S. McGaugh. 
	"Modified Newtonian Dynamics (MOND): Observational Phenomenology and Rlativistic Extensions",
	Living Rev. Relativity, 15, 10 (2012).
\bibitem{entropic} E.P. Verlinde, 
	"On the Origin of Gravity and the Laws of Newton". 
	JHEP. {\bf 2011}, 29 (2011); arXiv:hep-th/1001.0785. 
\bibitem{gans8}  S, Ganesh,  European Phys. J. C, {\bf 84}, 935 (2024). 	arXiv:gr-qc/2403.13019 (2024).
"Many body gravity and the galaxy rotation curves",
\bibitem{gans8E}  S, Ganesh,  European Phys. J. C, {\bf 85}, 257 (2025).
"Errata to: Many body gravity and the galaxy rotation curves"
\bibitem{gans9}  S, Ganesh,  
"Many body gravity and the bullet cluster",
Astroparticle Physics, {\bf 167}, 103080 (2025); arXiv:gr-qc/2501.05126 (2025).

\bibitem{bullet1} Tucker, W.; Blanco, P.; Rappoport, S. (March 1998). David, L.; Fabricant, D.; Falco, E. E.; Forman, W.; Dressler, A.; Ramella, M. 
"1E 0657-56: A Contender for the Hottest Known Cluster of Galaxies". 
Astrophysical Journal Letters. 496 (1): L5. arXiv:astro-ph/9801120. 


\bibitem{bullet2} Clowe, Douglas; Gonzalez, Anthony; Markevich, Maxim (2004). 
"Weak lensing mass reconstruction of the interacting cluster 1E0657-558: Direct evidence for the existence of dark matter",
	Astrophys. J. 604 (2): 596–603. arXiv:astro-ph/0312273. 

\bibitem{bullet3} M. Markevitch; A. H. Gonzalez; D. Clowe; A. Vikhlinin; L. David; W. Forman; C. Jones; S. Murray and W. Tucker (2004). 
"Direct constraints on the dark matter self-interaction cross-section from the merging galaxy cluster 1E0657-56",
Astrophys. J. 606 (2): 819–824. arXiv:astro-ph/0309303. 

\bibitem{bullet4} Clowe, Douglas; et al. (2006). 
"A Direct Empirical Proof of the Existence of Dark Matter", 
	The Astrophysical Journal Letters. 648 (2): L109–L113. arXiv:astro-ph/0608407. 
\bibitem{gans7}  S, Ganesh,  
"5D thermal field theory, Einstein field equations and spontaneous symmetry breaking",
	Class. Quantum Grav. {\bf 40}, 045008 (2023). 	arXiv:hep-th/2301.04827 (2023).
\bibitem{gans6}  S. Ganesh ,  
	"Quantum theory, thermal gradients and the curved Euclidean space",
	Int. J. Mod. Phys. A,, 
	{\bf 37}, 17, 2250125 (2022);
	arXiv:hep-th/2206.13324 (2022).
\bibitem{gans5}  S. Ganesh and M. Mishra,  
	"The effect of temperature gradient on the heavy quark–antiquark potential using a gravity dual model",
	Progress of Theoretical and Experimental Physics, {\bf 2021}, 1, 013B09 (2021).

\bibitem{alan} Alan, Guth
	"Inflationary universe: A possible solution to the horizon and flatness problems, Phys. Rev. D, 23, 347, (1981).

\bibitem{potential1}  A.A. Starobinsky, "A new type of isotropic cosmological models without singularity", Physics Letters, {\bf 91B}, 1 (1980).
\bibitem{potential2}  J. Martin, C.Ringeval, V. Vennin "Encyclopaedia Inflationaris", Phys Dark Univ, 5-6, 75-235 (2014).

\bibitem{graviton} Gia Dvali, Cesar Gomez, "Black Hole's Quantum N-Portrait" arXiv:hep-th/1112.3359 (2011).







\bibitem{PnS} Michael E. Peskin, Daniel V. Schroeder, "An introduction to Quantum Field theory", Westview Press (1995). 

\bibitem{tfr} Tully, R. B., Fisher, J. R., 
	"A New Method of Determining Distances to Galaxies",
	Astronomy and Astrophysics. {\bf 54}, 3, 661–673 (1977). 


\bibitem{wbs3} Indranil Banik et al. 
	"Strong constraints on the gravitational law from Gaia DR3 wide binaries",
	Monthly Notices of the Royal Astronomical Society, {\bf 527}, 3, (2024),




\bibitem{rot_lens} Tristan Faber and Matt Visser  
	"Combining rotation curves and gravitational lensing: how to measure the equation of state of dark matter in the galactic halo".
	Monthly Notices of the Royal Astronomical Society, {\bf 372}, 136-142  (2006),
\bibitem{ngc6505} C.M.O'Riordan et. al. 
	"Euclid: A complete Einstein ring in NGC 6505", 
	Astronomy and Astrophysics, {\bf 694}, A145 (2025).

\bibitem{sofue_mw} Y. Sofue, 
	"Grand Rotataion Curve and Dark-Matter Halo in the Mily Way Galaxy",
	Astronomical Society of Japan, {\bf 64}, 75 (2012).

\bibitem{hammer} Anna-Christina Eilers1,2, David W. Hogg1,3,4,5, Hans-Walter Rix1, and Melissa K. Ness,
"The Circular Velocity Curve of the Milky Way from 5 to 25 kpc",
	The Astrophysical Journal, {\bf 871}, 1 (2019); arXiv:astro-ph.GA/1810.09466 (2018).



\end{thebibliography}
\end{document}